\documentclass[aps,prd,tightenlines,floats,twocolumn,nofootinbib,amsmath,amssymb,showpacs,superscriptaddress]{revtex4}
\usepackage{amsmath,amssymb,amsfonts}
\usepackage{graphicx}
\usepackage{enumerate} 
\usepackage{colordvi} 
\usepackage{bm}

\newcommand{\bfv}{\mbox{\boldmath$v$}}

\newcommand{\bfx}{\mbox{\boldmath$x$}}

\newcommand{\bfk}{\mbox{\boldmath$k$}}
\newcommand{\bfp}{\mbox{\boldmath$p$}}
\newcommand{\bfq}{\mbox{\boldmath$q$}}

\newcommand{\bfOmg}{\mbox{\boldmath$\Omega$}}

\newcommand{\sigmav}{\sigma_{\rm d}}

\newcommand{\aap}{Astron. \& Astrophys.}
\newcommand{\physrep}{Phys. Rep.}
\newcommand{\RegPT}{{\texttt{RegPT}}}
\newcommand{\RegPTfast}{{\texttt{RegPT-fast}}}
\newcommand{\mptbreeze}{{\texttt{mptbreeze}}}
\begin{document}
\title{{\RegPT}:  Direct and fast calculation of regularized cosmological 
power spectrum at two-loop order}
\author{Atsushi Taruya}
\affiliation{Research Center for the Early Universe, School of Science, 
The University of Tokyo, Bunkyo-ku, Tokyo 113-0033, Japan}
\affiliation{
Kavli Institute for the Physics and Mathematics of the Universe, Todai Institutes for Advanced Study, the University of Tokyo, Kashiwa, Chiba 277-8583, Japan (Kavli IPMU, WPI)}
\author{Francis Bernardeau}
\affiliation{Institut de Physique Th\'eorique, CEA, IPhT, F-91191
  Gif-sur-Yvette, France.\\
  \ \ CNRS, URA 2306, F-91191, Gif-sur-Yvette, France}
\author{Takahiro Nishimichi}
\affiliation{
Kavli Institute for the Physics and Mathematics of the Universe, Todai Institutes for Advanced Study, the University of Tokyo, Kashiwa, Chiba 277-8583, Japan (Kavli IPMU, WPI)}
\author{Sandrine Codis}
\affiliation{Institut d'Astrophysique de Paris, 98 bis boulevard Arago, 
75014 Paris, France}
\bigskip
\date{\today}
%
\begin{abstract}
We present a specific prescription for the calculation of cosmological 
power spectra, exploited here at two-loop order 
in perturbation theory (PT), based on the multi-point propagator expansion. 
In this approach power spectra are 
constructed from the regularized expressions of the propagators that 
reproduce both the resummed behavior in the
high-$k$ limit and the standard  PT results at low-$k$. With the help of 
$N$-body simulations, 
we show that such a construction gives robust and accurate predictions 
for both the density power spectrum and 
the correlation function  at percent-level  in the weakly non-linear regime. 
We then present an algorithm that allows accelerated evaluations of
all the required diagrams by reducing the computational tasks to 
one-dimensional integrals. This is achieved by means of pre-computed kernel 
sets defined for appropriately chosen fiducial models. The computational 
time for two-loop results is then reduced from a few minutes, with the 
direct method, to a few seconds with the fast one. The robustness and 
applicability of this method are tested against 
the power spectrum {\tt cosmic emulator} from which a wide variety of 
cosmological models can be explored. The fortran program with which direct 
and fast calculations of power spectra can be done, 
\RegPT, is publicly released as part of this paper. 
\end{abstract}

\maketitle

\maketitle

\section{Introduction}
\label{sec:intro}
Since recombination, the large-scale structure of the Universe has evolved 
dominantly under the influence of 
both the cosmic expansion and the force of gravity acting on a pressure-less 
fluid. 
The statistical nature of its spatial clustering is therefore expected to 
bring valuable cosmological information about 
the dynamics of the cosmic expansion and structure formation. 
Of particular importance is the measurement of baryon acoustic oscillations 
(BAOs) imprinted on the power spectrum or two-point correlation function 
(e.g., \cite{Eisenstein:2005su,Percival:2009xn,Blake:2011wn,
Seo:2012xy,Anderson:2012sa}) from which one can precisely determine 
the cosmological distance to the high-redshift universe, and henceforth 
clarify the nature of late-time cosmic acceleration 
(e.g., \cite{Seo:2003pu,Blake:2003rh,Glazebrook:2005mb,Shoji:2008xn,
Padmanabhan:2008ag}). Precious  information regarding the growth of 
structure are and will also be obtained from redshift-space 
distortions (e.g., \cite{Linder:2007nu,Guzzo:2008ac,Yamamoto:2008gr,
Song:2008qt,Blake:2011rj}) and weak lensing measurements (see 
\cite{2000A&A...358...30V,2008A&A...479....9F} and review papers 
\cite{2001PhR...340..291B,2003astro.ph..5089V}) at scales ranging to the 
linear or quasi-linear to the non-linear regimes. 
This could be captured with unprecedented details  with the ongoing and 
future surveys, thanks to their redshift depth and large angular area, such 
as the Sloan Digital Sky Survey III\footnote{{\tt www.sdss3.org}}, the WiggleZ
survey\footnote{{\tt wigglez.swin.edu.au}}, 
the Subaru Measurement of Imaging and 
Redsfhits\footnote{{\tt sumire.ipmu.jp/en/}}, the 
Dark Energy Survey\footnote{{\tt www.darkenergysurvey.org}}, 
the BigBOSS project\footnote{{\tt bigboss.lbl.gov/index.html}}, 
the Physics of 
the Accelerating Universe collaboration\footnote{{\tt www.pausurvey.org}} and 
the ESA/Euclid survey \footnote{{\tt www.euclid-ec.org}}.

With the  advent of such wealth of observations,  there is therefore 
a growing interest in the development of theoretical tools to accurately 
compute  the statistical quantities of the large-scale structure.
At decreasing redshift and scale, the evolution of the large-scale structure 
however deviates significantly 
from the linear theory prediction and  non-linear gravitational clustering 
effects have to be taken into account. While $N$-body simulations can be 
relied upon in specific cases, because of the range of scales
to be covered and the variety of models to explore, they should be 
complemented by analytical investigations that aim at
computing the statistical properties of the large-scale structure from first 
principles, henceforth extending the validity range of linear calculations. 
It is to be noted that even at the scale of BAOs, linear calculations and 
one-loop standard PT corrections perform poorly (see e.g., 
\cite{Crocce:2005xy,Carlson:2009it,Taruya:2009ir}) asking for more advanced 
PT calculations. The improvement of perturbation theory is thus a critical 
issue for the scientific exploitation of the coming surveys. 
Various  resummation schemes have been proposed in Refs.
\cite{Crocce:2005xy,Crocce:2005xz,Crocce:2007dt,Matsubara:2007wj, 
Matsubara:2008wx,McDonald:2006hf,Izumi:2007su,Taruya:2007xy, Taruya:2009ir, 
Pietroni:2008jx,Matarrese:2007wc,Valageas:2003gm, Valageas:2006bi} 
that aim at improving upon standard schemes. The aim of this paper is not 
to compare them but to propose, and test, a specific scheme that can be used 
routinely in practice.

In this paper, we are particularly interested in one of the resummation 
treatments, advocated in Ref.~\cite{Bernardeau:2008fa}. In this approach, 
the standard PT expansion is re-organized by introducing the multi-point 
propagators. These are the ensemble average of the infinitesimal variation of 
the cosmic fields with respect to the initial conditions. A key property 
shown in the previous reference is that
all the statistical quantities such as power spectra and bispectra can be 
re-constructed by an expansion series written solely in terms of the 
multi-point propagators. This is referred to as the multi-point propagator 
expansion or $\Gamma$-expansion. 
The advantage of this approach is that the non-perturbative properties, 
which can be obtained in standard PT by summing up infinite series of PT 
expansions, are whole encapsulated in the multi-point propagators, including 
the effect of 
vertex renormalization. Furthermore, the $\Gamma$-expansion has been found 
to be valid not only 
for Gaussian initial conditions, but also  for non-Gaussian 
ones~\cite{Bernardeau:2010md}. 
The construction of accurate calculation scheme for power spectra and bispectra 
can then be split in pieces that can be tested separately.

The second key property that leads us to consider such objects is that their 
global shape, e.g. their whole $k$-dependence, can be computed in a 
perturbation theory context and compared to $N$-body results thanks to the 
high-$k$ exponential damping tail they all exhibit 
\cite{Bernardeau:2008fa,Bernardeau:2011vy}. 
All these properties make the multi-point propagators the most important 
building blocks in the $\Gamma$-expansion
and the focus of our modeling efforts. In the following we will in particular 
make full use of the novel regularization scheme proposed in 
Ref.~\cite{Bernardeau:2011dp} that allows to 
consistently  interpolate between  standard PT results at low-$k$ and the 
expected resummed behavior at high-$k$.
This scheme has been explicitly tested for the two-point propagators up to 
two-loop order in  Ref.~\cite{Bernardeauetal2012a}
and for (specific shapes of) the three-point propagators in 
Ref.~\cite{Bernardeau:2011dp}.

The first objective of this paper is to present an explicit 
calculation of the non-linear power spectrum and 
correlation function of the cosmic density field based on this regularized 
treatment. Of particular interest is the extent to which the proposed scheme 
for $\Gamma$-expansion works beyond standard PT when corrections at 
next-to-next-to-leading, i.e. two-loop, order 
are included. Results will be checked with $N$-body simulations. 
We will see that the $\Gamma$-expansion with the regularized treatment of 
propagators, which we hereafter call \RegPT, has good convergence
properties and agree remarkably well with simulations entirely covering the 
scales of BAOs at any redshift. 


The  second objective of this paper is to design and exploit a method to 
accelerate the power spectrum computations. Power spectra calculations in 
the context of  \RegPT~calculations are rather involved requiring 
multi-dimensional integrations that have to be done with time-consuming 
Monte Carlo calculations. Typically, computing the power spectrum at percent 
level from our scheme takes several minutes. While this is acceptable when a 
handful of models have to be computed, this is an obstacle when a large 
domain of parameter space has to be systematically explored. 
Making use of the $\Gamma$-expansion functional form, we found though that
it is possible to exploit a novel technique for accelerated calculation, 
in which only one-dimensional integrals need to be evaluated
while ensuring the same precision as rigorous \RegPT~calculations. 
The bottom line of this approach is to see the resulting nonlinear 
power spectrum as a functional of the linear power spectrum and then Taylor 
expand this form with respect to the linear spectrum shape.  We found that 
for well chosen fiducial models, it is sufficient to Taylor expand to
first order only.  We are then led to prepare in advance a set of kernel 
functions encoding the \RegPT~results for  well chosen fiducial models, 
whose normalizations are left floating, 
from which the \RegPT~predictions for the target model can be calculated. 
We publicly release the fortran code, \RegPT, as a part of this 
paper\footnote{
The code is available at \\
{\footnotesize{\tt www-utap.phys.s.u-tokyo.ac.jp/\~\,ataruya/regpt\_code.html}}
}.

The organization of this paper is as follows. 
We begin by recalling the basic equations for cosmic fluid 
and perturbation theory in Sec.~\ref{sec:RegPT}.
We introduce the multi-point propagator and give the power spectrum 
expression based on the $\Gamma$-expansion. With the regularized treatment 
of multi-point propagators, in Sec.~\ref{sec:power_spec}, we examine the 
the power spectrum calculations including the 
corrections up to the two-loop order, and investigate their
UV and IR sensitivity in evaluating the PT kernels.  
Then, in Sec.~\ref{sec:comparison}, a detailed comparison between 
PT calculation and $N$-body simulation is presented, and the accuracy and 
range of validity of PT calculation is checked. Based on this, 
Sec.~\ref{sec:PTreconst} describes in detail the method to accelerate the 
power spectrum calculations. 
Robustness and applicability of the accelerated \RegPT~calculations to 
a wide range of cosmological models are tested 
against power spectrum {\tt cosmic emulator} code in 
Sec.~\ref{subsec:reliable}.   Finally, in Sec.~\ref{sec:conclusion},
we conclude and explore practical extensions of this work. 
The description of the publicly released code, \RegPT, is presented in 
Appendix~\ref{appendix:RegPT_code}.

\section{Equations of motion and the $\Gamma$-expansion}
\label{sec:RegPT}

\subsection{Equations of motion}
\label{subsec:EoM}

In what follows, we consider the 
evolution of cold dark matter (CDM) plus baryon systems neglecting 
the tiny fraction of (massive) neutrinos. 
Owing to the single-stream approximation of the collisionless Boltzmann
equation, which is thought to be quite accurate an approximation 
on large scales, the evolution of 
the CDM plus baryon system can be treated as an irrotational and 
pressure-less fluid system whose governing equations are continuity 
and Euler equations in addition to the Poisson equation (see 
Ref.~\cite{Bernardeau:2001qr} for review). In the Fourier 
representation, these equations are further reduced to a 
more compact form. Let us introduce the two-component multiplet 
(e.g.,\cite{Crocce:2005xy}):  
\begin{align}
\Psi_a(\bfk;t)=\Bigl(\delta(\bfk;t), \,\,
-\frac{\theta(\bfk;t)}{f(t)} \Bigr),  
\end{align}
where the subscript $a=1,\,2$ selects the density and the velocity 
components of CDM plus baryons, with $\delta$ and 
$\theta(\bfx)\equiv\nabla\cdot \bfv(\bfx)/(a\,H)$, where 
$a$ and $H$ are the scale factor of the Universe and the Hubble 
parameter, respectively.    
The function $f(t)$ is given by $f(t)\equiv d\ln D(t)/d\ln a$, and 
the quantity $D(t)$ being the linear growth factor. 
Then, in terms of the new time variable 
$\eta\equiv\ln D(t)$, the evolution equation for the vector quantity 
$\Psi_a(\bfk;t)$ becomes 
\begin{align}
&\left[\delta_{ab}\,\frac{\partial}{\partial \eta}+\Omega_{ab}(\eta)\right]
\Psi_b(\bfk;\eta)
\nonumber\\
&\qquad=\int\frac{d^3\bfk_1\,d^3\bfk_2}{(2\pi)^3}
\delta_D(\bfk-\bfk_1-\bfk_2)\,\gamma_{abc}(\bfk_1,\bfk_2)\,
\nonumber\\
&\qquad\times\Psi_b(\bfk_1;\eta)\,\Psi_c(\bfk_2;\eta),
\label{eq:vec_fluid_eq}
\end{align}
where we used the summation convention, that is the repetition of the same 
subscripts  indicates the sum over the whole multiplet components. 
In the above, the quantity
$\delta_D$ is the Dirac delta function,  
and the time-dependent matrix $\Omega_{ab}(\eta)$ is given by
\begin{equation}
\Omega_{ab}(\eta)=\left(
\begin{array}{cc}
{\displaystyle 0} & {\displaystyle -1 }
\\
{\displaystyle -\frac{3}{2f^2}\Omega_{\rm m}(\eta)} \quad&\quad 
{\displaystyle \frac{3}{2f^2}\Omega_{\rm m}(\eta)-1}
\end{array}
\right) 
\label{eq:matrix_M}
\end{equation}
with the quantity $\Omega_{\rm m}(\eta)$ being the density parameter of 
CDM plus baryons at a given time. The vertex function 
$\gamma_{abc}$ becomes
\begin{eqnarray}
\gamma_{abc}(\bfk_1,\bfk_2) =
\left\{
\begin{array}{ccl} 
\frac{1}{2}\left\{1+\frac{\bfk_2\cdot\bfk_1}{|\bfk_2|^2}\right\} &;& 
(a,b,c)=(1,1,2) 
\\
\\
\frac{1}{2}\left\{1+\frac{\bfk_1\cdot\bfk_2}{|\bfk_1|^2}\right\} &;& 
(a,b,c)=(1,2,1) 
\\
\\
\frac{(\bfk_1\cdot\bfk_2)|\bfk_1+\bfk_2|^2}{2|\bfk_1|^2|\bfk_2|^2} &;& 
(a,b,c)=(2,2,2) 
\\
\\
     0                       &;&  \mbox{otherwise.}
\end{array}
\right.
\label{eq:def_Gamma}
\end{eqnarray}
Eq.~(\ref{eq:vec_fluid_eq}) can be recast as the integral equation   
(e.g., \cite{Bernardeau:2001qr,Crocce:2005xy}) 
\begin{align}
&\Psi_a(\bfk;\eta)=g_{ab}(\eta,\eta_0)\,\phi_b(\bfk) +
\int_{\eta_0}^{\eta} d\eta' g_{ab}(\eta,\eta')\,
\nonumber\\
&\quad\times
\int\frac{d^3\bfk_1\,d^3\bfk_2}{(2\pi)^3}\delta_D(\bfk-\bfk_1-\bfk_2)\,
\gamma_{bcd}(\bfk_1,\bfk_2)
\nonumber\\
&\quad\qquad\times\Psi_c(\bfk_1;\eta')\Psi_d(\bfk_2;\eta').
\label{eq:formal_sol}
\end{align}
The quantity $\phi_a(\bfk)\equiv\Psi_a(\bfk,\eta_0)$ 
denotes the initial condition, and the function $g_{ab}$ denotes 
the linear propagator satisfying the following equation,
\begin{align}
\left[\delta_{ab}\frac{\partial}{\partial\eta}+\Omega_{ab}(\eta)\right]
g_{bc}(\eta,\eta')=0, 
\label{eq:linear_prop}
\end{align}
with the boundary condition $g_{ab}(\eta,\eta)=\delta_{ab}$. 
The statistical properties of the field $\Psi_a$ are encoded in the 
initial field $\phi_a$, for which we assume Gaussian statistics.  
The power spectrum of $\phi_a$ is defined as 
\begin{align}
\langle\phi_a(\bfk)\phi_b(\bfk')\rangle = 
(2\pi)^3\,\delta_D(\bfk+\bfk')\,P_{ab,0}(k). 
\label{eq:def_P_0}
\end{align}
In what follows, most of the calculations will be made assuming the 
contribution of decaying modes of linear perturbation can be neglected. 
This implies that the field $\phi_a(\bfk)$ is factorized as 
$\phi_a(\bfk)=\delta_0(\bfk) u_a$ with $u_a=(1,1)$, and 
thus the initial power spectrum is written as $P_{ab,0}(k)=P_0(k)u_au_b$.

\begin{figure}


\includegraphics[width=8cm,angle=0]{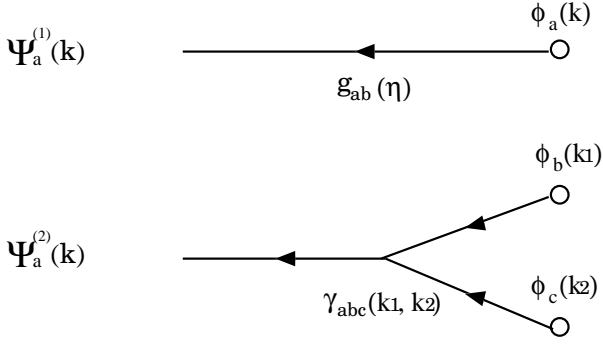}

\vspace*{0.0cm}

\caption{Diagrammatic representation of the standard PT expansion.  
\label{fig:diagram_SPT}}
\end{figure}

Using the formal expression (\ref{eq:formal_sol}), a perturbative solution 
is obtained by expanding the fields in terms of the initial fields:  
\begin{align}
\Psi_a(\bfk;\eta) = \sum_{n=1}^{\infty} \,\Psi^{(n)}_a(\bfk;\eta).
\label{eq:PT_expansion}
\end{align}
The expression of the solution at each order is written as 
\begin{align}
&\Psi_a^{(n)}(\bfk;\eta) = \int\frac{d^3\bfk_1\cdots d^3\bfk_n}{(2\pi)^{3(n-1)}}
\delta_D(\bfk-\bfk_1-\cdots-\bfk_n)
\nonumber\\
&\quad\qquad\times\mathcal{F}_a^{(n)}
(\bfk_1,\bfk_2,\cdots,\bfk_n;\eta)\,\delta_0(\bfk_1)\cdots\delta_0(\bfk_n).
\label{eq:PT_solution}
\end{align}
The kernel $\mathcal{F}_a^{(n)}$ is generally a complicated time-dependent 
function, but can be constructed in terms of the quantities $\gamma_{abc}$ and 
$g_{ab}$. Examples of the solutions are shown diagrammatically in 
Fig.~\ref{fig:diagram_SPT}. Because we are interested in the late-time 
evolution of large-scale structure only, we can take the limit 
$\eta_0\to-\infty$. As a consequence, the fastest growing term is the 
only surviving one and the kernel is simplified into
\begin{align}
\mathcal{F}_a^{(n)}(\bfk_1,\cdots,\bfk_n)=
e^{n\,\eta}\,F_{a,{\rm sym}}^{(n)}(\bfk_1,\cdots,\bfk_n),
\label{eq:PT_kernel}
\end{align}
where the function $F_{a,{\rm sym}}^{(n)}$ is the symmetrized standard PT kernel, 
sometimes written as $F_{a,{\rm sym}}^{(n)}=(F_n,\,\,G_n)$, 
whose explicit expressions are obtained from recursion relations as recalled in~\cite{Bernardeau:2001qr}. 

\subsection{$\Gamma$-expansion and regularized PT treatment}
\label{subsec:Gamma-expansion}

\begin{figure*}[ht]


\includegraphics[width=8cm,angle=0]{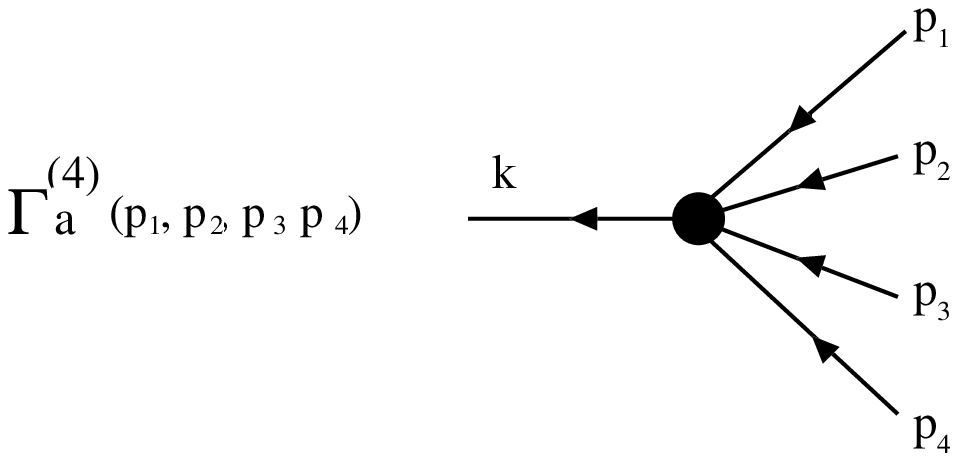}

\vspace*{0.0cm}

\caption{
Example of the multi-point propagator, $\Gamma_a^{(4)}$. A large
filled circle symbolically represents all possible contributions that 
enter into the fully non-linear propagator. 
A part of those contributions can be seen graphically 
using PT expansion (see Figs.~\ref{fig:diagram_Gamma2spt} 
and \ref{fig:diagram_Gamma2reg} for three-point propagator $\Gamma_a^{(2)}$). 
\label{fig:diagram_Gamma4}}


\vspace*{0.5cm}

\includegraphics[width=18cm,angle=0]{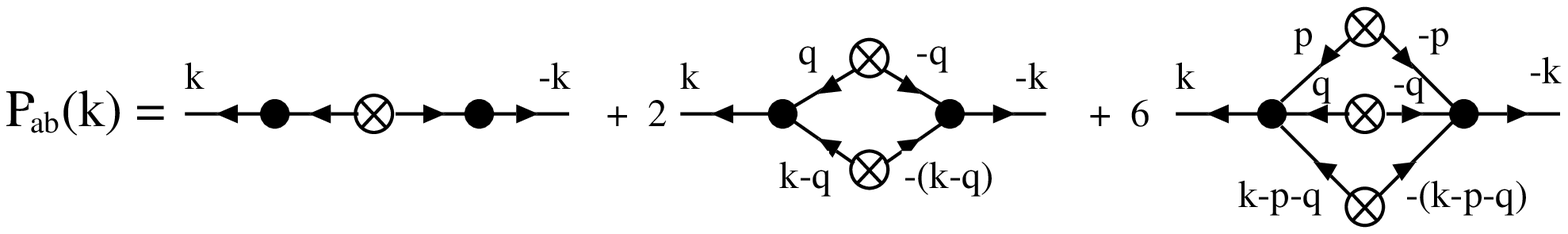}

\vspace*{0.0cm}

\caption{Diagrammatic representation of the power spectrum by means of 
  $\Gamma$-expansion. Here, the result up to the two-loop order is shown. 
  In each contribution of the diagrams, 
  the multi-point propagators are ``glued'' together at the crossed circles where
  the initial power spectra $P_0(k)$ are inserted. 
\label{fig:diagram_pk}}
\end{figure*}

In this paper, we are more specifically interested in the power spectra 
$P_{ab}(k;\eta)$, defined as  
\begin{equation}
\Bigl\langle\Psi_a(\bfk;\eta)\Psi_b(\bfk';\eta)\Bigr\rangle=
(2\pi)^3\,\delta_{\rm D}(\bfk+\bfk')\,P_{ab}(|\bfk|;\eta). 
\label{eq:def_Pk} 
\end{equation}
Substituting a set of perturbative solutions (\ref{eq:PT_solution}) 
into the above definition,  
it is straightforward to obtain the successive perturbative expressions
for the power spectra. This is the standard PT treatment where the
initial fields values are seen as the perturbative variables.
The standard PT calculations have, however, been shown to produce 
ill-behaved higher-order corrections that lack good convergence properties. 

As an alternative 
to the standard PT framework, it has been recently advocated 
by many authors that the PT expansion can be re-organized by introducing 
non-perturbative quantities to improve the resulting convergence of 
the expansion. The $\Gamma$-expansion is one such
non-perturbative framework, and the so-called 
multi-point propagators constitute the 
building blocks of this $\Gamma$-expansion. 
Denoting the $(p+1)$-point propagator by $\Gamma^{(p)}$, we define 
\begin{align}
&\frac{1}{p!}\left\langle
\frac{\delta^p\Psi_a(\bfk,\eta)}{\delta\phi_{c_1}(\bfk_1)
\cdots \delta\phi_{c_p}(\bfk_p)}\right\rangle =\delta_{\rm D}
(\bfk-\bfk_{1\cdots p})
\nonumber\\
&\quad\times\frac{1}{(2\pi)^{3(p-1)}} \,
\Gamma_{a c_1\cdots c_p}^{(p)}(\bfk_1,\cdots,\bfk_p;\eta).  
\end{align}
With these objects, the power spectra is shown to be expressed as 
\cite{Bernardeau:2008fa},
\begin{align}
&P_{ab}(|\bfk|;\eta) = 
\sum_{t=1}^{\infty}
t!\int\frac{d^3\bfq_1\cdots d^3\bfq_t}
{(2\pi)^{3(t-1)}}\,\,\delta_{\rm D}(\bfk-\bfq_{1\cdots t})\,
\nonumber\\
&\times\Gamma_a^{(t)}(\bfq_1,\cdots,\bfq_t;\eta)
\Gamma_b^{(t)}(\bfq_1,\cdots,\bfq_t;\eta)\,P_0(q_1)\cdots P_0(q_t), 
\label{eq:G_expansion_Pk}
\end{align}
where we introduced the shorthand notation,
\begin{align}
\Gamma_a^{(t)}(\bfq_1,\cdots,\bfq_t;\eta)=
\Gamma_{ac_1\cdots c_t}^{(t)}(\bfq_1,\cdots,\bfq_t;\eta)u_{c_1}\cdots
u_{c_t}. 
\end{align}
The diagrammatic representation for multi-point propagator and  
the power spectrum is respectively 
shown in Figs.~\ref{fig:diagram_Gamma4} and \ref{fig:diagram_pk}.

The construction of the $\Gamma$-expansion 
is rather transparent, and 
like Eq.~(\ref{eq:G_expansion_Pk}), one easily finds the expressions for 
the higher-order statistical quantities such as bispectrum. 
Another important point is that one can 
exploit the asymptotic properties of the propagators $\Gamma^{(p)}$ beyond 
perturbation theory expansions. To be precise, in the high-$k$ limit,
higher-order contributions can be systematically computed at all orders, and 
as a result of summing up all the contributions,  the multi-point propagators 
are shown to be exponentially suppressed~\cite{{Bernardeau:2008fa,Bernardeau:2011vy}},
\begin{align}
&\Gamma^{(p)}_a(\bfk_1,\cdots,\bfk_p;\eta)
\nonumber\\
&\qquad
\stackrel{k\to\infty}{\longrightarrow} 
\exp\left\{-\frac{k^2\sigmav^2e^{2\eta}}{2}\right\}
\Gamma^{(p)}_{a,{\rm tree}}(\bfk_1,\cdots,\bfk_p;\eta)
\label{eq:Gamma_high-k}
\end{align}
with $k=|\bfk_1+\cdots+\bfk_p|$. This is the generalization of the 
result for the two-point propagator in Ref.~\cite{Crocce:2005xz}. 
Here, the quantity $\Gamma^{(p)}_{a,{\rm tree}}$ is the lowest-order 
non-vanishing propagator obtained from the standard PT calculation, and 
$\sigmav$ is the one-dimensional root-mean-square of the displacement 
field defined by,
\begin{align}
\sigmav^2=\frac{1}{3}\int\frac{d^3\bfq}{(2\pi)^3}\,\frac{P_0(q)}{q^2}.
\label{eq:sigam_v}
\end{align}

\begin{figure*}[ht]

\vspace*{-0.5cm}

\includegraphics[width=18cm,angle=0]{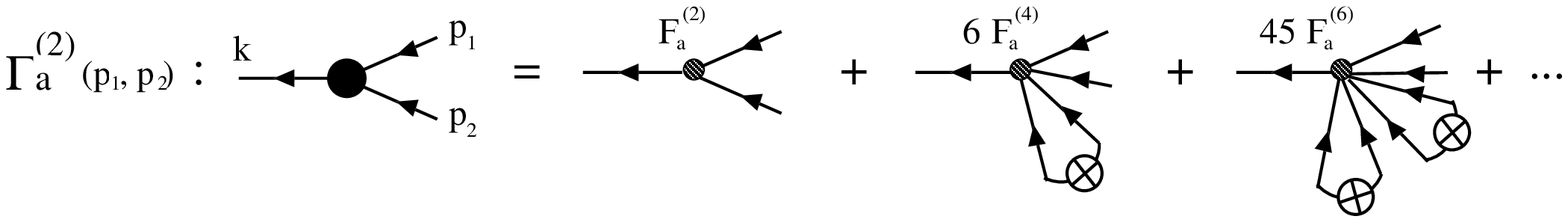}

\vspace*{0.0cm}

\caption{Diagrammatic representation of the standard PT expansion 
for three-point propagator, $\Gamma_a^{(2)}$. For fastest growing-mode
contribution, the standard PT kernels, $F_{a,{\rm sym}}^{(n)}$,  
form the basic pieces of PT expansion, 
depicted as incoming lines connected to a single outgoing line at 
the shaded circle. In the case of $\Gamma_a^{(2)}$, the leading-order 
contribution is $F_{a,{\rm sym}}^{(2)}$, and successively 
the kernels $F_{a,{\rm sym}}^{(4)}$ and $F_{a,{\rm sym}}^{(6)}$ appear as 
higher-order contributions, for which pairs of the incoming lines are 
glued at the crossed circle, which indicates the initial power 
spectrum $P_0$, forming closed loops. 
\label{fig:diagram_Gamma2spt}}


\includegraphics[width=16cm,angle=0]{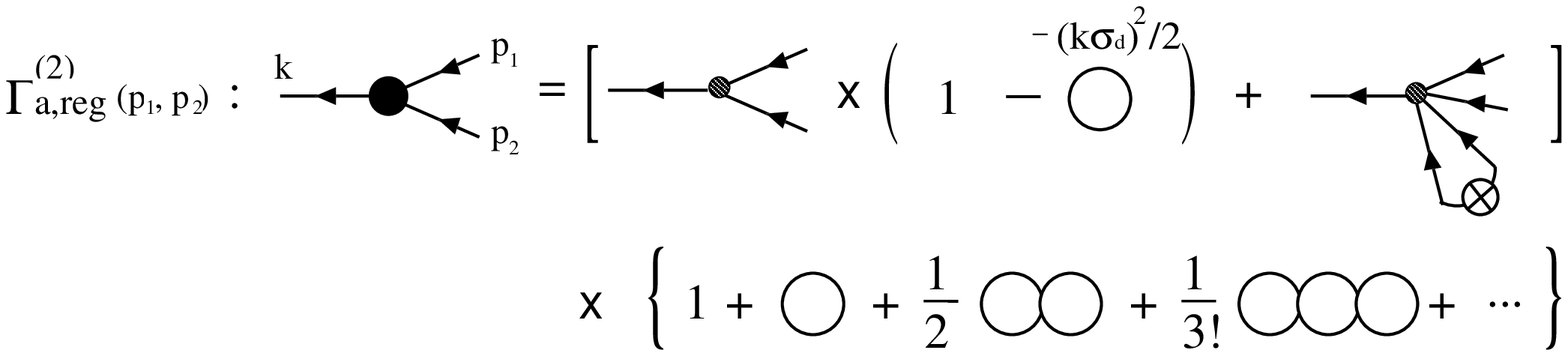}

\vspace*{0.0cm}

\caption{Diagrammatic representation of the regularized 
three-point propagator, $\Gamma_{a,{\rm reg}}^{(2)}$. In the high-$k$ limit, 
the higher-loop contribution for three-point propagator 
behaves like Eq.~(\ref{eq:Gamma}), indicating that each 
loop diagram ($\Gamma^{(2)}_{a,n\mbox{-}{\rm loop}}$) is effectively 
split into tree diagram ($F_{a,{\rm sym}}^{(2)}$) 
and self-loop diagram ($\{-(k\sigmav)^2/2\}^n/n!$), the latter of which 
is depicted as open loops. Systematically summing up all the 
higher-loop contributions, we recover Eq.~(\ref{eq:Gamma_high-k}),  
which is graphically shown as the infinite sum of 
open-loop diagrams in the brace. To reproduce the standard PT result 
at low-$k$,  the one-loop diagram is inserted in the bracket, 
and the tree diagram is multiplied by the counter term 
$\{1+(k\sigmav)^2/2\}$. 
\label{fig:diagram_Gamma2reg}}
\end{figure*}

The form of Eq. (\ref{eq:Gamma_high-k}) does not however provide a good description of the
propagators at all scale. At low-$k$ the propagators are expected to approach their standard PT expressions
that can be written formally,
\begin{align}
&\Gamma_a^{(p)}(\bfk_1,\cdots,\bfk_p;\eta)
=\Gamma_{a,{\rm tree}}^{(p)}(\bfk_1,\cdots,\bfk_p;\eta)
\nonumber\\
&\qquad\qquad\qquad\qquad
+\sum_{n=1}^\infty\Gamma_{a,n\mbox{-}{\rm loop}}^{(p)}(\bfk_1,\cdots,\bfk_p;\eta).
\label{eq:Gamma_loop_expand}
\end{align}
For the dominant growing-mode contribution we are interested in, 
each correction term is expressed in terms of the standard PT kernels as,
\begin{align}
\Gamma^{(p)}_{a,{\rm tree}}(\bfk_1,\cdots,\bfk_p;\eta)&=
e^{p\,\eta}\,F_{a,{\rm sym}}^{(p)}(\bfk_1,c\dots,\bfk_p),
\label{eq:Gamma-p_tree}
\end{align}
for the tree-level contribution, and 
\begin{widetext}
\begin{align}
\Gamma_{a,n\mbox{-}{\rm loop}}^{(p)}
(\bfk_1,\cdots,\bfk_p;\eta)
&=e^{(2n+p)\,\eta}\,c_n^{(p)}
\int\frac{d^3\bfp_1\cdots d^3\bfp_n}{(2\pi)^{3n}}\,
F_{a,{\rm sym}}^{(2n+p)}(\bfp_1,-\bfp_1,\cdots,\bfp_n,-\bfp_n,\bfk_1,\cdots,\bfk_p)
P_0(p_1)\cdots P_0(p_n)
\nonumber\\
&\equiv e^{(2n+p)\,\eta}\,\,
\overline{\Gamma}_{a,n\mbox{-}{\rm loop}}^{(p)}(\bfk_1,\cdots,\bfk_p)
\label{eq:Gamma-p_nloop}
\end{align}
\end{widetext}
for the $n$-loop order contributions, where the coefficient $c_n^{(p)}$ is given by 
$c_n^{(p)}=\,_{(2n+p)}C_p\ (2n-1)!!$ where 
is the binomial coefficient. The graphical representation of 
the standard PT expansion is shown in Fig.~\ref{fig:diagram_Gamma2spt}. 
The important remark in Eq.~(\ref{eq:Gamma-p_nloop}) 
is that each perturbative correction possesses the following asymptotic form,
\begin{align}
\Gamma^{(p)}_{a,n\mbox{-}{\rm loop}}
\stackrel{k\to\infty}{\longrightarrow} 
\frac{1}{n!} \left(-\frac{k^2\sigmav^2e^{2\eta}}{2}\right)^{n}
\Gamma^{(p)}_{a,n\mbox{-}{\rm tree}}, 
\label{eq:Gamma}
\end{align}
which consistently recovers the expression (\ref{eq:Gamma_high-k}) 
when we sum up all the loop contributions. This indicates the existence of a  matching scheme which smoothly interpolates 
between the low-$k$ and high-$k$ results for any multi-point propagator. Such a scheme has been proposed in  
Ref~\cite{Bernardeau:2011dp} where a novel regularized 
scheme, in which the low- and high-$k$ behaviors are jointly reproduced, is derived.  
The construction of the {\it regularized} propagator is totally unambiguous. They can incorporate 
an arbitrary number of loop corrections.

Restricting the results to the growing mode contributions, 
the {\it regularized} propagators are expressed in a transparent way 
in terms of the standard PT results, and one gets 
\begin{widetext}
\begin{align}
\Gamma_{a,{\rm reg}}^{(p)}(\bfk_1,\cdots,\bfk_p;\eta)=e^{p\,\eta}\left[
F^{(p)}_{a,{\rm sym}}(\bfk_1,\cdots,\bfk_p)\left\{1+
\frac{k^2\sigmav^2e^{2\eta}}{2}\right\}+
e^{2\eta}\,\overline{\Gamma}^{(p)}_{a,{\rm 1\mbox{-}loop}}(\bfk_1,
\cdots,\bfk_p)\right]
\exp\left\{-\frac{k^2\sigmav^2e^{2\eta}}{2}\right\}, 
\end{align}
\end{widetext}
which consistently reproduces one-loop PT results at low-$k$. 
An example
of the regularized propagator valid at one-loop order is diagrammatically 
shown in Fig.~\ref{fig:diagram_Gamma2reg}.  
This construction is easily generalized to include the higher-order
PT corrections at low-$k$. For instance, 
the regularized propagator including the corrections 
up to the two-loop order becomes,
\begin{widetext}
\begin{align}
&\Gamma_{a,{\rm reg}}^{(p)}(\bfk_1,\cdots,\bfk_p;\eta)=e^{p,\eta}\left[
F^{(p)}_{a,{\rm tree}}(\bfk_1,\cdots,\bfk_p)\left\{1+
\frac{k^2\sigmav^2e^{2\eta}}{2}+
\frac{1}{2}\left(\frac{k^2\sigmav^2e^{2\eta}}{2}\right)^2\right\}
\right.
\nonumber\\
&\left.\qquad\qquad\qquad+\,
e^{2\eta}\,\overline{\Gamma}^{(p)}_{\rm 1\mbox{-}loop}(\bfk_1,\cdots,\bfk_p)
\left\{1+\frac{k^2\sigmav^2e^{2\eta}}{2}\right\}+e^{4\eta}\,
\overline{\Gamma}^{(p)}_{\rm 2\mbox{-}loop}(\bfk_1,\cdots,\bfk_p)\right]
\exp\left\{-\frac{k^2\sigmav^2e^{2\eta}}{2}\right\}.
\end{align}
\end{widetext}
Note that the functions $\overline{\Gamma}^{(p)}_{n\mbox{-}{\rm loop}}$ are the 
scale-dependent part of the propagator 
defined by Eq.~(\ref{eq:Gamma-p_nloop}).

\section{Power spectrum calculation from regularized $\Gamma$-expansion}
\label{sec:power_spec}

\subsection{Power spectrum at two-loop order}
\label{subsec:expression}

Since the proposed regularized propagators 
preserve the expected low-$k$ and high-$k$ behaviors, 
the convergence of the $\Gamma$-expansion adopting the regularization scheme 
would be much better than the standard PT expansion. In this paper, 
applying this regularized PT treatment, we will explicitly demonstrate 
the power spectrum calculations at two-loop order, and comparing 
those predictions
with $N$-body simulations, the validity and precision of PT treatment are 
discussed. From Eq.~(\ref{eq:G_expansion_Pk}), 
the explicit expression for the power spectrum valid up to 
the two-loop order becomes,
\begin{widetext}
\begin{align}
&P_{ab}(k;\eta) = 
\Gamma_{a,{\rm reg}}^{(1)}(k;\eta)\Gamma_{b,{\rm reg}}^{(1)}(k;\eta)P_0(k)+
2\int\frac{d^3\bfq}{(2\pi)^3}\,
\Gamma_{a,{\rm reg}}^{(2)}(\bfq,\bfk-\bfq;\eta)
\Gamma_{b,{\rm reg}}^{(2)}(\bfq,\bfk-\bfq;\eta)P_0(q)P_0(|\bfk-\bfq|)
\nonumber\\
&\qquad\qquad\qquad\qquad+6\int\frac{d^6\bfp d^3\bfq}{(2\pi)^6}\,
\Gamma_{a,{\rm reg}}^{(3)}(\bfp,\bfq,\bfk-\bfp-\bfq;\eta)
\Gamma_{b,{\rm reg}}^{(3)}(\bfp,\bfq,\bfk-\bfp-\bfq;\eta)
P_0(p)P_0(q)P_0(|\bfk-\bfp-\bfq|)
\label{eq:pk_Gamma_reg_2loop}
\end{align}
with the regularized propagators given by
\begin{align}
&\Gamma_{a,{\rm reg}}^{(1)}(k;\eta)=e^{\eta}\left[
1+
\frac{k^2\sigmav^2e^{2\eta}}{2}+
\frac{1}{2}\left(\frac{k^2\sigmav^2e^{2\eta}}{2}\right)^2
+e^{2\eta}\,\overline{\Gamma}^{(1)}_{a,{\rm 1\mbox{-}loop}}(k)
\left\{1+\frac{k^2\sigmav^2e^{2\eta}}{2}\right\}+
e^{4\eta}\,\overline{\Gamma}^{(1)}_{a,{\rm 2\mbox{-}loop}}(k)\right]
\exp\left\{-\frac{k^2\sigmav^2e^{2\eta}}{2}\right\},
\label{eq:Gamma1_reg}
\\
& \Gamma_{a,{\rm reg}}^{(2)}(\bfq,\bfk-\bfq;\eta)=e^{2\eta}\left[
F_{a,{\rm sym}}^{(2)}(\bfq,\bfk-\bfq)\left\{1+
\frac{k^2\sigmav^2e^{2\eta}}{2}\right\}+
e^{2\eta}\,\overline{\Gamma}^{(2)}_{a,{\rm 1\mbox{-}loop}}(\bfq,
\bfk-\bfq)\right]
\exp\left\{-\frac{k^2\sigmav^2e^{2\eta}}{2}\right\},
\label{eq:Gamma2_reg}
\\
&\Gamma_{a,{\rm reg}}^{(3)}(\bfp,\bfq,\bfk-\bfp-\bfq;\eta)=
e^{3\eta}\,F_{a,{\rm sym}}^{(3)}(\bfp,\bfq,\bfk-\bfp-\bfq)
\exp\left\{-\frac{k^2\sigmav^2e^{2\eta}}{2}\right\}. 
\label{eq:Gamma3_reg}
\end{align}
\end{widetext}
Note that the higher-order contributions up to the two- and one-loop order
of the propagators are respectively included in the expression 
of the regularized propagators $\Gamma_{a,{\rm reg}}^{(1)}$ and 
$\Gamma_{a,{\rm reg}}^{(2)}$,  consistently with the 
$\Gamma$-expansion at two-loop order.

The power spectrum expression involves many integrals, but,  
most of them are reduced to two- or three-dimensional integrals if 
one uses the analytic expressions for the kernels of higher-loop 
corrections $\overline\Gamma$ in the 
regularized propagator. We use the expression in 
Ref.~\cite{Bernardeauetal2012a} 
to evaluate $\overline\Gamma^{(2)}_{\rm 1\mbox{-}loop}$, and adopt the 
fitting functions for the kernel of $\overline\Gamma^{(1)}_{\rm 2\mbox{-}loop}$ 
(see \cite{Bernardeauetal2012a}).  
We then apply the method of Gaussian quadrature to the 
numerical evaluation of the low-dimensional integrals. A bit cumbersome is 
the integral containing $\Gamma^{(3)}$. 
While it can be reduced to
a four-dimensional integral in principle, 
the expression of the resulting kernel would be very cumbersome 
and might not be suited for practical calculation. 
We thus adopt the Monte Carlo technique
of quasi-random sampling using the CUBA library \cite{Hahn:2004fe}, and  
evaluate the five-dimensional integral directly\footnote{Since the 
final result of the integration is expressed as a function of 
only the wavenumber $k$, the integrand possesses an azimuthal symmetry 
with respect to the vector, $\bfk$, indicating that the integral is reduced 
to a five-dimensional integral.}.

\begin{figure*}[t]

\vspace*{-2.0cm}

\includegraphics[width=10cm,angle=0]{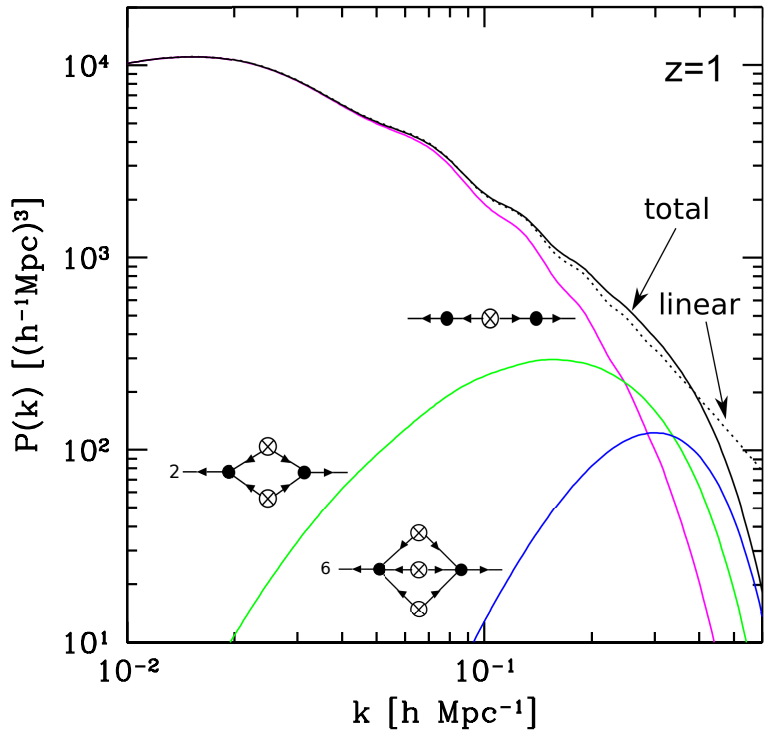}


\caption{Contribution of multi-point propagators to the power spectrum, 
$P(k)=P_{11}(k)$ at $z=1$.  
Magenta, green, and blue curves represent the 
power spectrum contributions 
from the first, second, and third terms at the right-hand-side 
of Eq.~(\ref{eq:pk_Gamma_reg_2loop}), respectively, 
each of which just corresponds to the diagram in Fig.~\ref{fig:diagram_pk}, 
involving $\Gamma_{\rm reg}^{(1)}$, 
$\Gamma_{\rm reg}^{(2)}$, and $\Gamma_{\rm reg}^{(3)}$.
Summing up these contributions, 
total power spectrum is shown in black solid line. 
For reference, linear power spectrum is also plotted as dotted line. 
\label{fig:pk_total}}
\end{figure*}

Fig.~\ref{fig:pk_total} illustrates 
an example how each correction 
term in the regularized $\Gamma$-expansion 
contributes to the total power spectrum. The plotted result is the 
density power spectrum, $P_{11}$, and 
the contribution of the term involving each multi-point propagator is 
separately shown. The three 
corrections contribute to the power spectrum at different scales,  
and the higher-order terms involving $\Gamma^{(2)}_{\rm reg}$ and 
$\Gamma^{(3)}_{\rm reg}$ are well-localized, each producing one bump. This is 
a clear manifestation of the result of the regularized PT treatment, and 
it resembles what the RPT calculations by Ref.~\cite{Crocce:2007dt} give.

In the next section, the results of the regularized $\Gamma$-expansion will be 
compared with $N$-body simulations. But, before doing that, we will give 
several remarks and comments 
on the computation of the power spectrum in the 
subsequent subsection.

\subsection{Effect of running UV cutoff for $\sigmav$} 
\label{subsec:running_sigmav}

\begin{figure*}[t]

\vspace*{-0.5cm}

\includegraphics[width=11cm,angle=0]{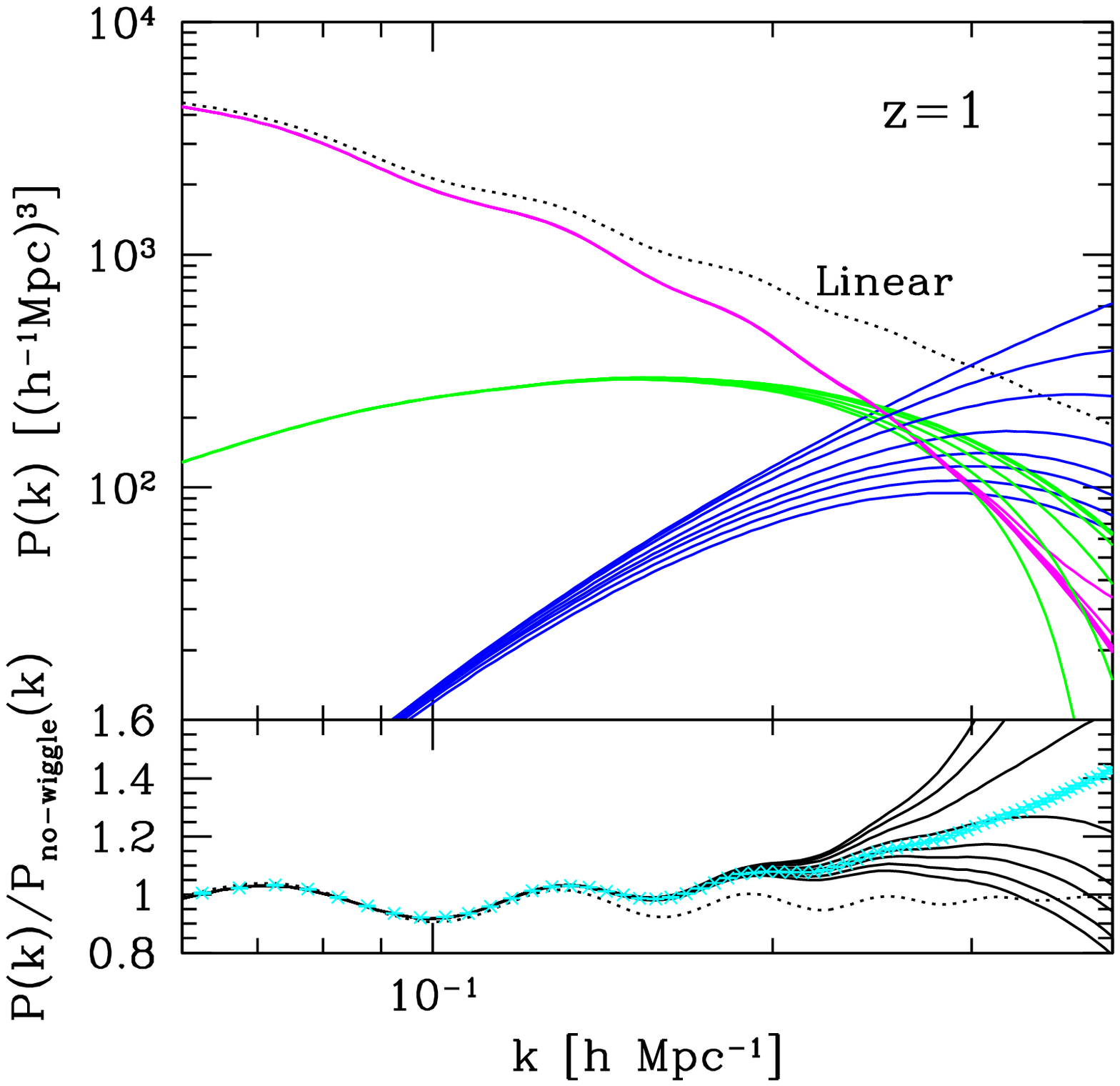}

\vspace*{-1.5cm}

\caption{Sensitivity of the power spectrum prediction at $z=1$ 
to the UV cutoff in the estimation of $\sigmav$. 
Top panel shows each contribution of the power spectrum corrections 
involving $\Gamma^{(1)}$ (magenta), $\Gamma^{(2)}$ (green), and 
$\Gamma^{(3)}$ (blue), 
respectively. Bottom panel shows the total sum of power spectrum divided 
by the smooth reference power spectrum, $P_{\rm no\mbox{-}wiggle}(k)$, which is
calculated from the no-wiggle formula of the linear transfer function 
in Ref.~\cite{Eisenstein:1997ik}. 
In each case, top lines represent the results obtained by setting $\sigmav=0$, 
while undermost lines show the cases adopting the value of $\sigmav$ 
without UV cutoff. 
The middle six lines represent the cases adopting the running UV cutoff
in estimating $\sigmav$, with cutoff $k_\Lambda(k)=k$, 
$k/2$, $k/3$, $k/5$, $k/10$, and $k/20$ (from 
bottom to top). As a reference, linear theory prediction is also plotted 
in both panels (dotted).  
\label{fig:pk_total_running_sigmav}}
\end{figure*}

Since the shape of the power spectrum given by 
Eq.~(\ref{eq:pk_Gamma_reg_2loop}) 
significantly depends on the exponential damping in the regularized 
propagators, we first comment on the effect of this 
function. As it has been shown, the exponential function arises from 
the summation of infinite series of perturbations at all order 
in the high-$k$ limit. Recently, Ref.~\cite{Bernardeau:2011vy} 
advocated that 
this exponential function can be interpreted as the result of resummation 
at hard part (high-$k$), and the displacement dispersion $\sigmav$ 
in the exponent must be evaluated in a consistent way that the domain of the 
integral is restricted to a soft part (low-$k$). This implies that 
depending on the scale of our interest, the boundary of the soft and 
hard domains can be changed, and the resulting quantity $\sigmav$ 
should be regarded as a scale-dependent function.

In Fig.~\ref{fig:pk_total_running_sigmav},
we examine the impact of 
the scale-dependent $\sigmav$ on the power spectrum at $z=1$. 
Plotted results are the contributions of the power spectrum corrections 
(upper) and the 
total power spectrum divided by the smooth reference linear spectrum 
(bottom). Here, we evaluate $\sigmav$ by introducing the running UV 
cutoff $k_\Lambda(k)$: 
\begin{align}
\sigmav^2(k)=\int_0^{k_\Lambda(k)}\frac{dq}{6\pi^2}\, P_0(q).
\label{eq:running_sigma_v}
\end{align}
Various curves in 
Fig.~\ref{fig:pk_total_running_sigmav}
represent the results with different prescription for the
running UV cutoff. 
The correction involving the four-point propagator 
$\Gamma_{\rm reg}^{(3)}$ is most sensitively affected by 
the running cutoff, and the resulting power spectrum significantly varies 
at scales $k\gtrsim0.2\,h$Mpc$^{-1}$. 
This is because the exponential damping 
manifests itself at the scale $k\sim  1/(e^{\eta}\sigmav)$ where 
the contribution from $\Gamma_{\rm reg}^{(3)}$, which contains no 
relevant terms counteracted with the exponential damping, 
becomes significant among the three corrections.

In the bottom panel of Fig.~\ref{fig:pk_total_running_sigmav},
we also plot 
the result of $N$-body simulations (see Sec.~\ref{subsec:simulation}). 
The comparison with simulation 
suggests that the PT calculation with running cutoff 
$k_\Lambda\sim k/2-k/5$ 
is favored, although there is no clear physical reason why this is so. Strictly speaking, the 
running IR cutoff might also be introduced in evaluating  
all the integrals in the power spectrum expression, so as to 
consistently discriminate between 
the contributions coming from soft and hard parts. 
Moreover, the running cutoff $k_\Lambda$ may also depend on the redshift. 
These complications mostly come from the ambiguity of the boundary between 
soft and hard domains in  our regularization scheme. 
For practical purpose to the cosmological application,
we postpone these issues to future investigation, and take a 
rather phenomenological approach. Hereafter, 
the running cutoff is only introduced in evaluating $\sigmav$,  
and we evaluate it according to 
Eq.~(\ref{eq:running_sigma_v}), setting the cutoff scale 
to $k_{\Lambda}=k/2$. With this treatment, we will see 
later that the PT prediction becomes improved compared to the 
standard PT calculation, and it reproduces the $N$-body results 
quite well at any redshift.

\begin{figure*}[t]

\vspace*{-0.5cm}

\includegraphics[width=8cm,angle=0]{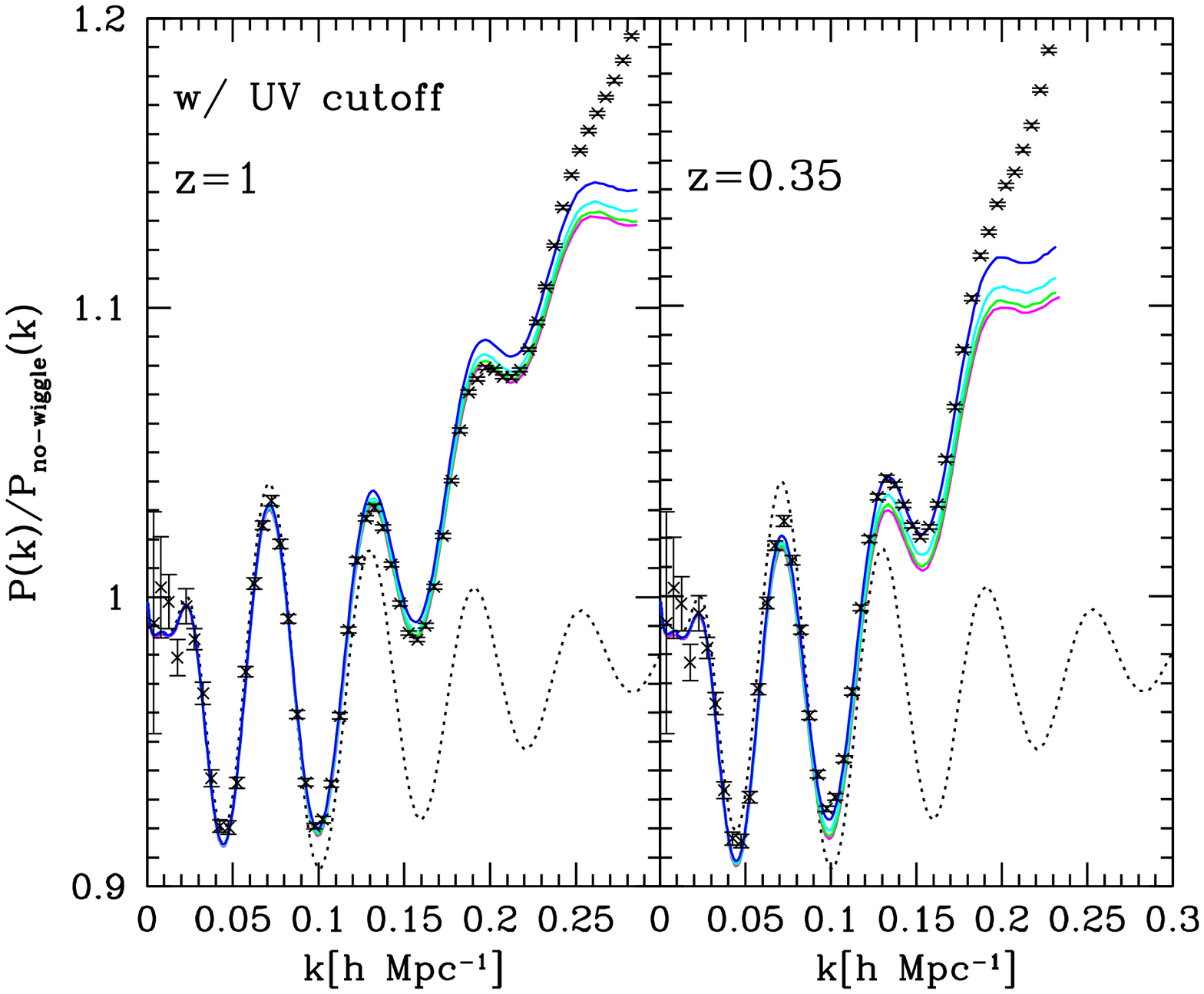}
\includegraphics[width=8cm,angle=0]{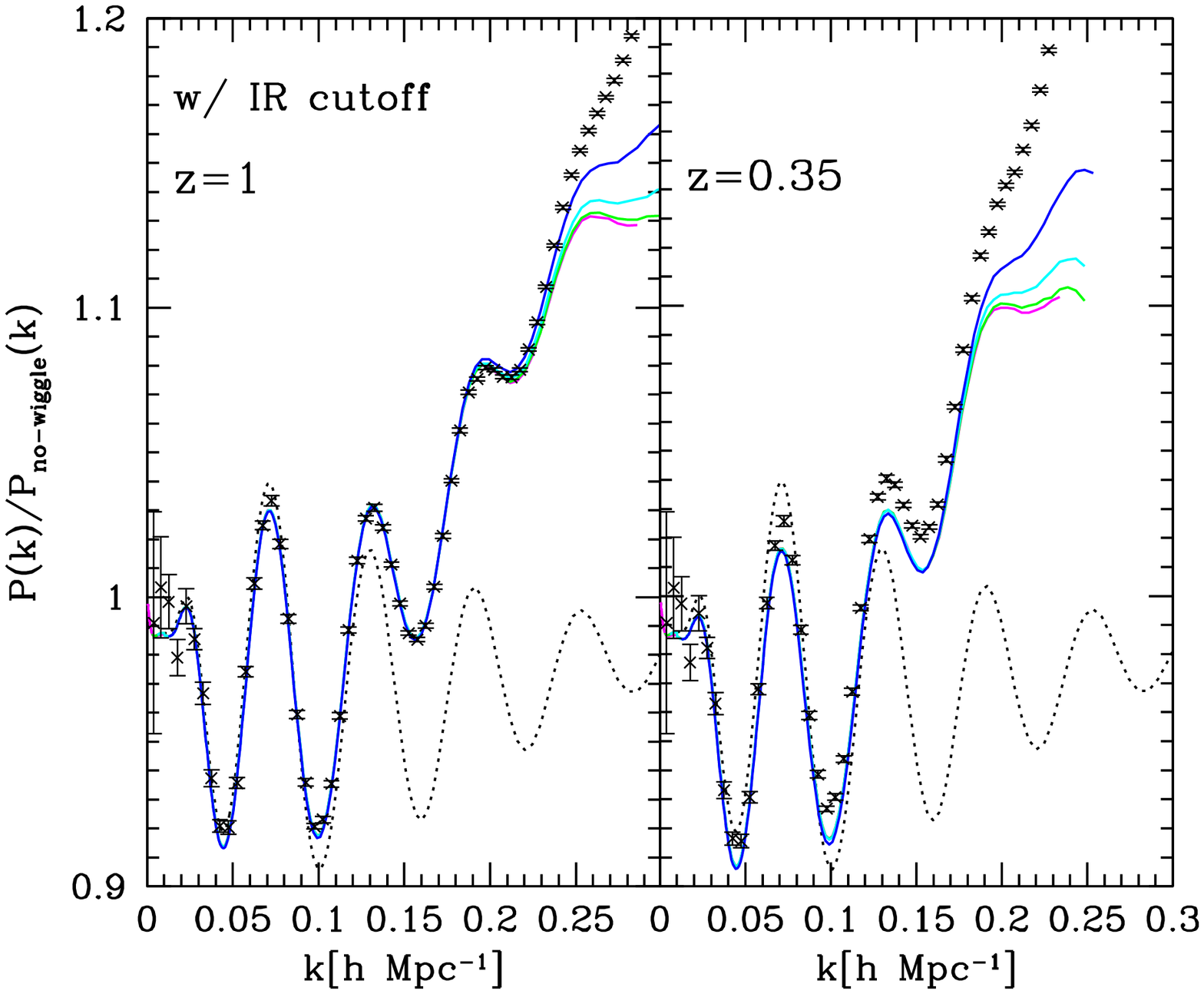}

\vspace*{0.0cm}

\caption{Sensitivity of the power spectrum prediction 
to the UV (left) and IR (right) cutoff. In each 
panel, the ratio of power spectrum,  
$P(k)/P_{\rm no\mbox{-}wiggle}(k)$, is plotted as function of $k$ 
at $z=1$ (left) and $0.35$ (right). In evaluating the integrals of 
the power spectrum corrections, the maximum wavenumber for the 
range of the integral is set to  
$2\pi$ (green), $\pi$ (cyan), and $\pi/2$ (blue) $h$\,Mpc$^{-1}$ in the 
left panel, while in the right panel, we change 
the minimum wavenumber $k_{\rm min}$ to 
$k_{\rm min}=2\pi/L_{\rm box}$ with $L_{\rm box}=2,048$ (green), $1,024$ 
(cyan), and $512\,h^{-1}$ Mpc (blue). The magenta lines indicate the 
results adopting the default set of parameters 
$(k_{\rm min},k_{\rm max})=(5\times10^{-4},10)\,h$ Mpc$^{-1}$.  
\label{fig:ratio_pk_kmax_kmin}}
\end{figure*}

\subsection{Sensitivity to IR and UV cutoff}
\label{subsec:IR_UV_sensitivity}

In computing the power spectrum, except for $\sigmav$, the domain of each
integral in Eq.~(\ref{eq:pk_Gamma_reg_2loop}) are usually taken 
broad enough so as to ensure the convergence of the results. 
In comparison with $N$-body simulations, however,   
a care must be taken because the available Fourier modes in 
simulations are restricted depending on the simulation box size
and/or mesh size of Fourier transform, which affect
both the efficiency of mode transfer and strength of mode coupling. 
The evolved result of the power spectrum would thus be changed, and 
it should be carefully compared with PT calculation, taking 
the finite resolution into consideration. 
Here, focusing on the BAO scales, 
we briefly discuss the sensitivity of the PT calculation to the IR and UV 
cutoff in the integrals. 

Fig.~\ref{fig:ratio_pk_kmax_kmin} shows the variation of 
the power spectra with respect to the UV (left) and IR (right) cutoff. 
In general, the kernel of integrals becomes broader for higher-loop 
corrections, and thus the two-point propagator $\Gamma^{(1)}_{\rm reg}$ 
containing the two-loop contribution is sensibly affected by
the UV cutoff. Note that the signs of one- and 
two-loop corrections in $\Gamma^{(1)}_{\rm reg}$ are opposite at BAO scales.   
Hence, as decreasing the cutoff wavenumber $k_{\rm max}$, 
the cancellation of each contribution is relatively relaxed, and 
the power spectrum amplitude gets increased. 
On the other hand, due to the lack of the long-wave modes,  
the IR cutoff not only decreases each contribution of the loop integrals, 
but also reduces $\sigmav$, leading to a slight suppression of 
the exponential damping. The net effect of the IR cutoff, especially at small 
scales $k\gtrsim0.2\,h\,$Mpc$^{-1}$, 
is that the latter overcomes the former, and the total 
power spectrum is slightly enhanced.

These results imply
that the effect of UV and IR 
cutoff not only affects the power spectrum shape at small scales,  
but also causes a slight offset in power spectrum amplitude 
at moderately large scales, $k\sim0.1\,h$Mpc$^{-1}$. 
The size of these effects is basically small, but 
would not be negligible in a percent-level comparison.  
Based on this remark, in what follows, we adopt the cutoff scales 
$(k_{\rm min},k_{\rm max})=(5\times10^{-4},10)\,h$ Mpc$^{-1}$ as 
default parameters to compute the power spectra. 
With this setup, \RegPT~calculation gives a mostly convergent 
result, which can be compared with high-resolution 
$N$-body simulations with a large box size. 

\vspace*{0.3cm}

\subsection{Comparison with MPTbreeze}
\label{subsec:MPTbreeze}

In Ref.~\cite{2012arXiv1207.1465C}, \mptbreeze, an alternative scheme has been 
proposed for the construction of
power spectra that is based on the same multi-point propagator expansion. 
This proposition is, however, based on 
simplified assumptions regarding the behavior of the multi-point propagators. 
More specifically, in \mptbreeze, the propagators
are assumed to take the form,
\begin{widetext}
\begin{equation}
\Gamma_{a,{\rm reg}}^{(p)}(\bfq_{1},\dots,\bfq_{p-1},\bfk-\bfq_{1 \cdots (p-1)};\eta)=
e^{p\,\eta}\,F_{a,{\rm sym}}^{(p)}(\bfq_{1},\dots,\bfq_{p-1},\bfk-\bfq_{1\cdots(p-1)})\,
\exp\left\{f_{a}(k)e^{2\eta}\right\},
\end{equation}
\end{widetext}
where $f_{1}(k)$ and $f_{2}(k)$ are the one-loop corrections to the density and velocity propagators, respectively. This form
corresponds to the late-time original expression of the exponentiation 
scheme initially put forward in Ref.~\cite{Crocce:2005xz}. It is shown in 
Refs.~\cite{Bernardeau:2011dp,Bernardeauetal2012a} that at one-loop order, 
this prescription gives nearly identical result for the two-point propagator 
to the prescription proposed in Ref.~\cite{Crocce:2005xz}. 
The \mptbreeze~prescription, however, ignores the impact of two-loop PT 
corrections on the two-point propagators. 
From the results presented in \cite{Bernardeauetal2012a}, it implies that 
\mptbreeze~might be outperformed by  \RegPT~at $z\gtrsim1$. 
On the other hand, the predictions of that scheme are made more robust 
because they are less sensitive to the UV 
part of the linear spectrum as discussed in that paper. 
Furthermore, the one-loop correction for the three- and four-point 
propagators are treated in an effective way. These simplified 
assumptions allow a more rapid calculations of the set of diagrams. 
It takes just a few seconds to get the expected shape in this
scheme. The computational time is, however, rather 
comparable to the fast implementation of \RegPT~ which we will present 
in Sec.~\ref{sec:PTreconst}.

\section{Comparison with $N$-body simulations}
\label{sec:comparison}

We are now in position to present quantitative comparisons between  \RegPT~calculations and $N$-body simulations. 
After briefly describing the $N$-body simulations in 
Sec.~\ref{subsec:simulation}, 
we show the results of power spectrum and two-point correlation 
function in Sec.~\ref{subsec:result} and \ref{subsec:result2}, respectively. 
Precision and validity of the PT predictions are discussed in detail.

\begin{figure*}[ht]
\hspace*{-1.2cm}
\includegraphics[width=9.5cm,angle=0]{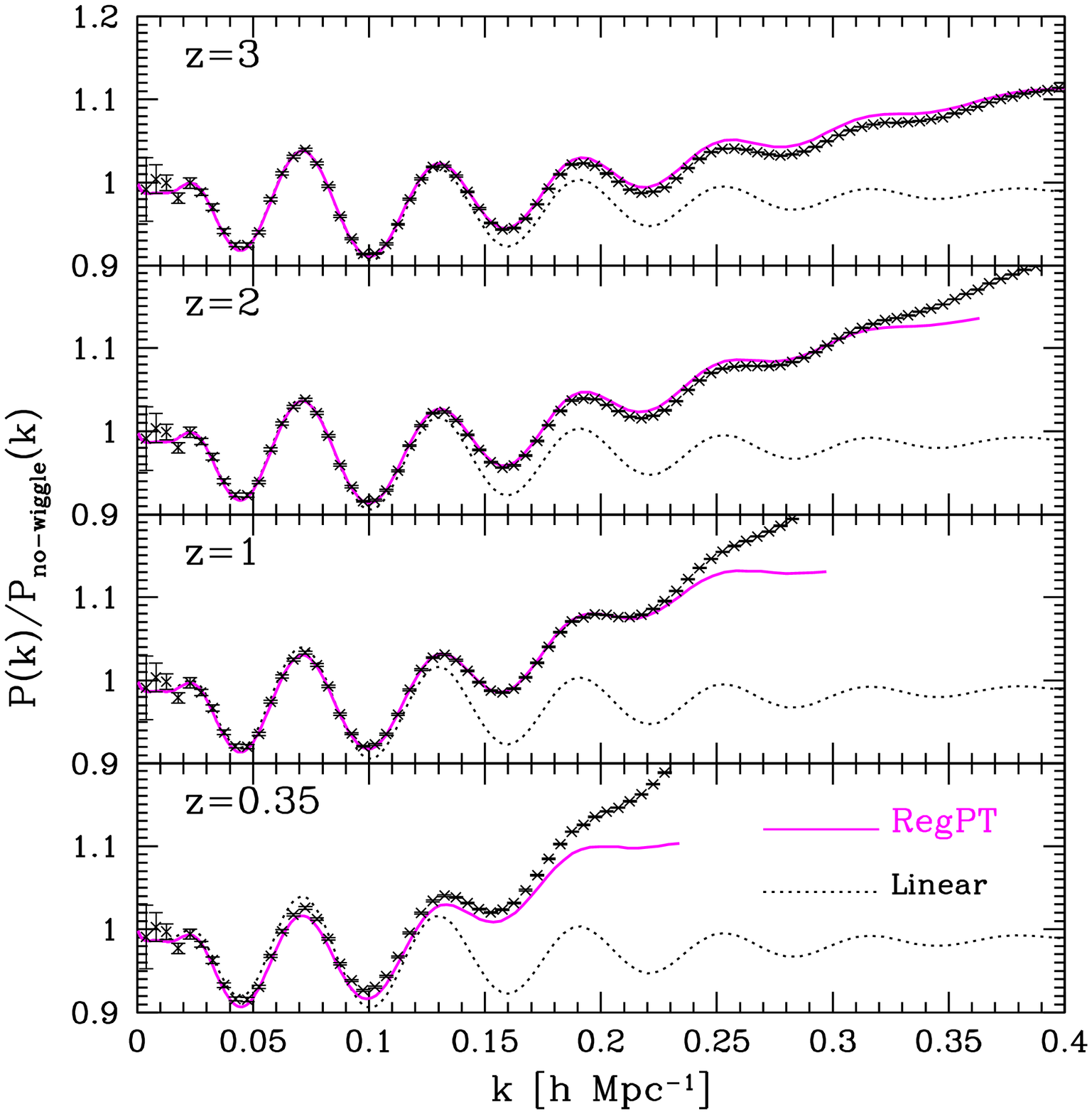}
\hspace*{-0.3cm}
\includegraphics[width=9.5cm,angle=0]{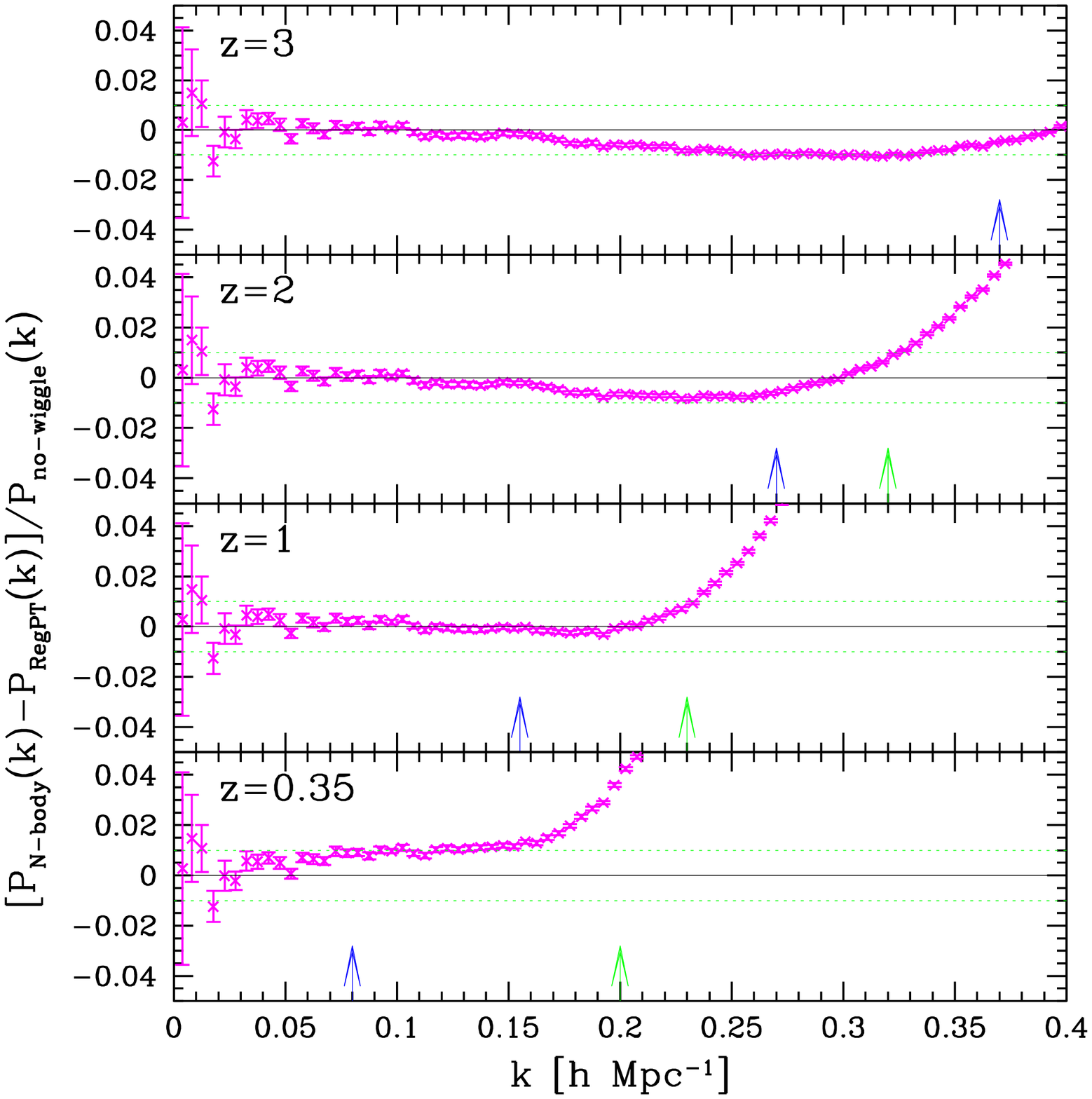}

\vspace*{0.5cm}

\caption{Comparison of power spectrum results between $N$-body 
simulations and \RegPT~calculations. In each panel, the results 
at $z=3$, $2$, $1$, and $0.35$ are shown (from top to bottom). 
Left panel shows the ratio of power spectrum to the smooth linear 
spectrum, $P(k)/P_{\rm no\mbox{-}wiggle}(k)$, where the reference spectrum 
$P_{\rm no\mbox{-}wiggle}(k)$ is calculated from the no-wiggle formula of 
the linear transfer function in Ref.~\cite{Eisenstein:1997ik}. 
Solid lines are the \RegPT~results, while dotted lines represent the linear 
theory predictions. Right panel plots the difference between 
$N$-body and \RegPT~results normalized by the no-wiggle spectrum, i.e., 
$[P_{\rm N\mbox{-}body}(k)-P_{\rm RegPT}(k)]/P_{\rm no\mbox{-}wiggle}(k)$. In each panel,
The vertical arrows respectively indicate the maximum wavenumber 
below which a percent-level agreement with $N$-body simulation is 
achieved with Lagrangian resummation theory 
\cite{Matsubara:2007wj,Okamura:2011nu} and 
closure theory \cite{Taruya:2009ir,Taruya:2007xy}, 
including the PT correction up to two-loop order. 
\label{fig:pkreal}}
\end{figure*}

\begin{figure*}[ht]
\hspace*{-1.0cm}
\includegraphics[width=10cm,angle=0]{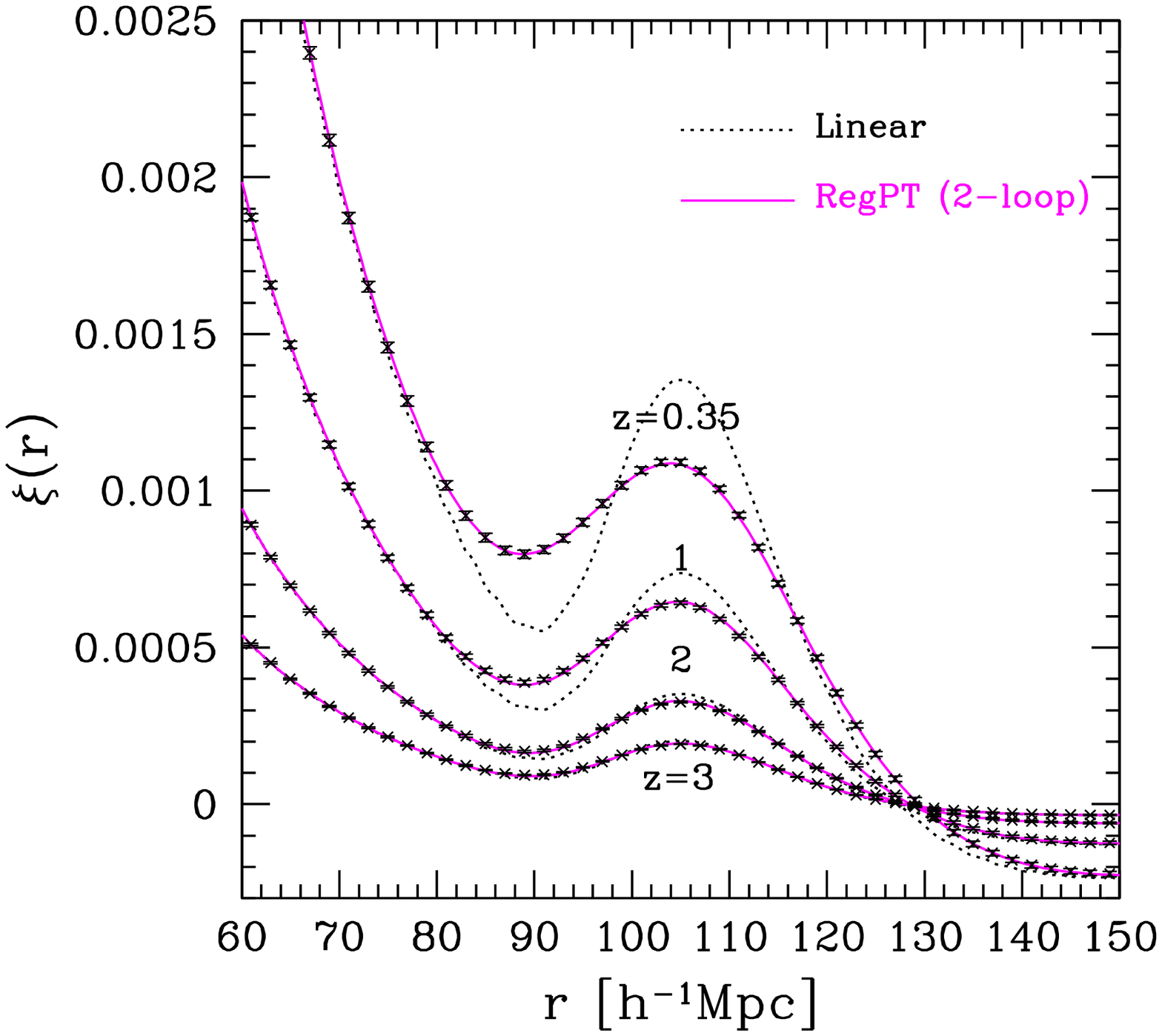}
\hspace*{-1.5cm}
\includegraphics[width=10cm,angle=0]{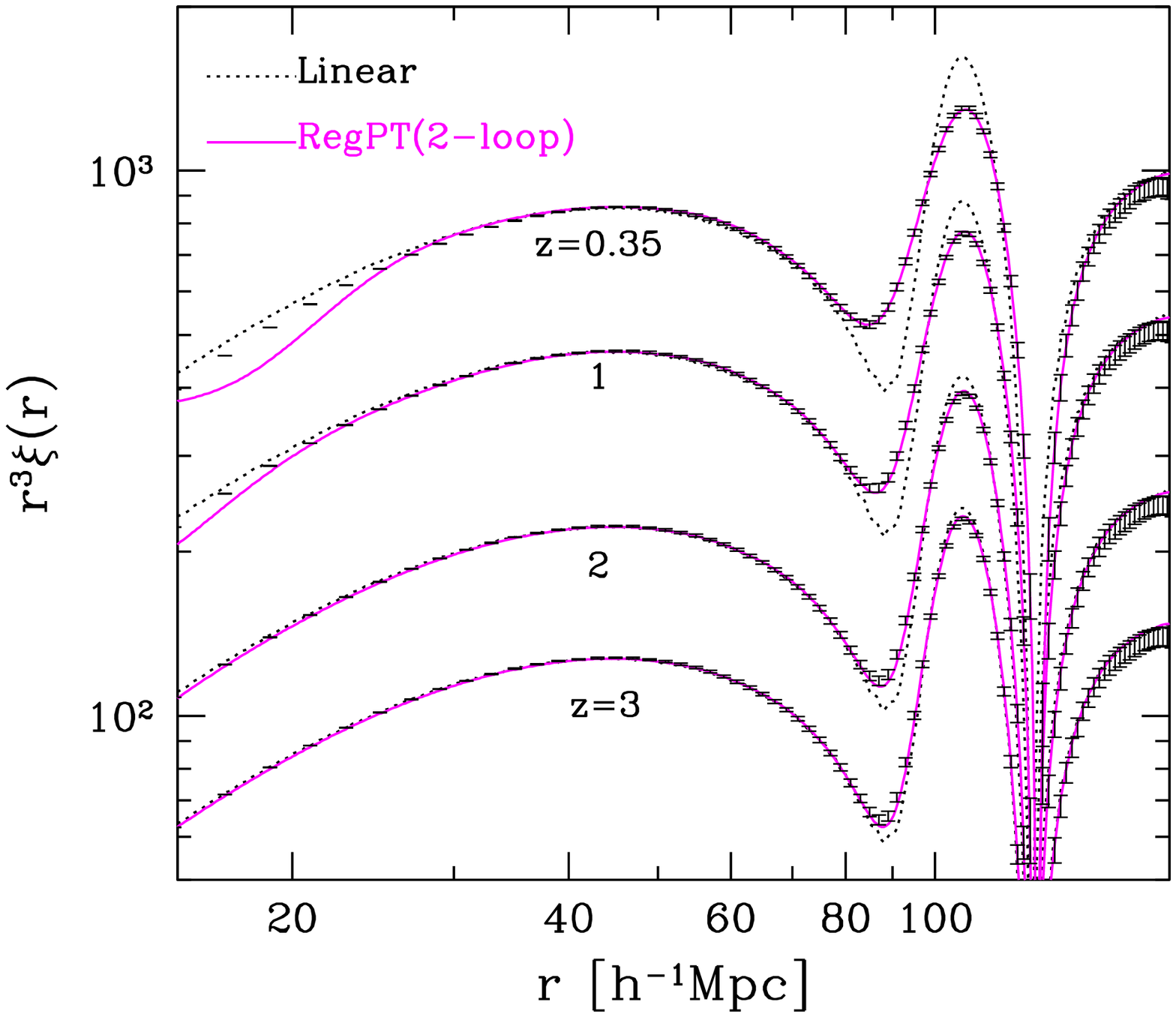}

\vspace*{-0.5cm}

\caption{Comparison of two-point correlation function between 
$N$-body and \RegPT~results at $z=3$, $2$, $1$, and $0.35$ (from 
bottom to top). In each panel, magenta solid, and black dotted lines
represent the prediction from \RegPT~and linear theory calculations, 
respectively. Left panel focuses on the behavior around baryon
acoustic peak in linear scales, while right panel shows 
the overall behavior in a wide range of separation in logarithmic 
scales.  Note that in right panel, the resulting 
correlation function is multiplied by the cube of the separation
for illustrative purpose. 
\label{fig:xireal}}
\end{figure*}

\begin{table*}[ht]
\caption{\label{tab:nbody_nishimichi} Cosmological parameters for $N$-body 
simulations ($\Lambda$CDM) }
\begin{ruledtabular}
\begin{tabular}{lcccc|ccccccc}
Name & $L_{\rm box}$ & \# of particles & $z_{\rm ini}$ 
& \# of runs & 
$\Omega_{\rm m}$ & $\Omega_{\Lambda}$ & $\Omega_{\rm b}/\Omega_{\rm m}$ & $w$ & $h$ & $n_s$ & $\sigma_8$ \\
\hline
\verb|wmap5| & $2,048h^{-1}$Mpc & $1,024^3$ & $15$ & $60$ & 
0.279 &  0.721 & 0.165 & -1 & 0.701 & 0.96 & $0.815_9$\\
\end{tabular}
\end{ruledtabular}
\end{table*}

\subsection{$N$-body simulations}
\label{subsec:simulation}

To compare \RegPT~calculations with $N$-body simulations, we ran a new set of
$N$-body simulations, 
which will be presented in more detail with an extensive convergence study in \cite{Nishimichi:inprep}.
This set of simulations can be regarded as an updated version of 
the one presented in \cite{Taruya:2009ir} with much larger total volume and a more careful setup 
to achieve a smaller statistical and systematic error. 
The data were created by a public $N$-body code \verb|GADGET2| 
\cite{Springel:2005mi} 
with cubic boxes of side length $2,048\,h^{-1}$Mpc, and $1,024^3$ particles. 
The cosmological parameters adopted in these $N$-body simulations are 
basically the same as in the previous one, and are determined 
by the five-year WMAP observations \cite{Komatsu:2008hk} 
(see Table~\ref{tab:nbody_nishimichi}). 
The initial conditions were generated by a parallelized version of 
the \verb|2LPT| code \cite{Crocce:2006ve}, developed in 
Ref.~\cite{Valageas:2010yw}. After several tests given in 
Ref.~\cite{Valageas:2010yw}, a lower initial redshift $z_{\rm init}$ turns 
out to give a more reliable estimate for the power spectrum at BAO scales,  
and we thus adopt the initial redshift $z_{\rm init}=15$. 
With this setup, we have created $60$ independent realizations and  
the data were stored at redshifts $z=3$, $2$, $1$, and $0.35$. The 
total volume at each output redshift is $515\,h^{-3}$Gpc$^3$, which is 
statistically sufficient for a detailed comparison with PT calculations.

We measure both the matter power spectrum and the 
correlation function. For the power spectrum, we adopt 
the Cloud-in-Cells interpolation, and construct 
the Fourier transform of the density field assigned on the $1,024^3$ grids. 
As for the estimation of the
two-point correlation function, we adopt the grid-based calculation using the 
Fast Fourier Transformation \cite{Taruya:2009ir}. 
Similarly to the power spectrum analysis, 
we first compute the square of the density field 
on each grid point in Fourier space. 
Then, applying the inverse Fourier 
transformation, we take the average over 
separation vectors and realizations,
and finally obtain the two-point correlation function. The 
implementation of this method, together with 
a convergence test, is presented in more detail in Ref~.\cite{Taruya:2009ir}.

\subsection{Power spectrum}
\label{subsec:result}

Let us first present the power spectrum results. 
Left panel of Fig.~\ref{fig:pkreal} shows the ratio of the power spectra, 
$P(k)/P_{\rm no\mbox{-}wiggle}(k)$, while the right panel plots 
the fractional difference between $N$-body simulations and PT 
calculations, defined by 
$[P_{\rm N\mbox{-}body}(k)-P_{\rm PT}(k)]/P_{\rm no\mbox{-}wiggle}(k)$
(where $P_{\rm no\mbox{-}wiggle}(k)$ is calculated from the no-wiggle formula of 
the linear transfer function in Ref.~\cite{Eisenstein:1997ik}). 
Overall, the agreement between \RegPT~and $N$-body simulations 
is remarkable at low-$k$, and a percent-level agreement is 
achieved up to a certain wavenumber. 
For a decreasing redshift, the non-linearities develop and the
applicable range of PT calculations inevitably becomes narrower, 
however, compared to the standard PT predictions, 
the \RegPT~result can reproduce the $N$-body trend 
over an even wider range. Indeed, the range of agreement with $N$-body 
simulations is rather comparable to other improved PTs 
including higher-order corrections, 
such as closure theory \cite{Taruya:2009ir,Taruya:2007xy}, 
and better than some of those predictions. 
For reference, we compute 
the power spectra from closure and Lagrangian resummation theory (LRT) 
\cite{Matsubara:2007wj,Okamura:2011nu} at two-loop order, 
and estimate the range of a percent-level 
agreement with $N$-body simulations, the results of which are   
respectively depicted as green and blue vertical arrows
in right panel of Fig.~\ref{fig:pkreal}. Note that at $z=3$, 
the range of agreement for closure theory exceeds 
the plotted range, and is not shown here.

Although the \RegPT~treatment gives a very good performance 
comparable to or even better than other improved PTs, 
a closer look at Fig.~\ref{fig:pkreal} reveals  several sub-percent 
discrepancies. 

\begin{itemize}
\item One is the low-$k$ behavior at $z=0.35$, which exhibits a small 
discrepancy with $N$-body simulation. Our investigations indicate that it is probably due to a poor convergence of standard PT expansion, since 
the low-$k$ behavior of regularized propagators 
heavily relies on the standard PT treatment. 
To be specific, the convergence of $\Gamma^{(1)}_{\rm reg}$ is the main
source of this discrepancy. Indeed, if $\Gamma^{(1)}_{\rm reg}$ is computed at one-loop order only, 
the power spectrum is enhanced, and then
N-body results at low-$k$ lie in between the two predictions.
The impact of the high order PT corrections to the two-point propagator are specifically studied
in a separate publication, \cite{Bernardeauetal2012a}.
\item
Another discrepancy can be
found in the high-$z$ results, which temporally overshoot 
the $N$-body results at mid-$k$ regime ($k\sim0.2$\,--\,$0.3\,h$\,Mpc$^{-1}$). 
It is unlikely to be due to a poor convergence of
standard PT expansion. We rather think that the performances of the $N$-body simulations might be responsible
for this (small) discrepancy. We have tested several runs 
with different resolutions, and found that the low-resolution simulation
with a small number of particles tends to underestimate the power at high-$z$. 
Possible reason for this comes from the precision of 
force calculation around the intervening scales, where the Tree and 
Particle-Mesh algorithms are switched, and is mainly attributed to 
the inaccuracy of the Tree algorithm. Though the intervening scale is 
usually set at a sufficiently small scale, with a low-resolution simulation,  
it may affect the large-scale dynamics with noticeable effects at higher 
redshifts. Systematic studies on 
the convergence and resolution of $N$-body simulations will be 
reported elsewhere \cite{Nishimichi:inprep}. 
\end{itemize}

\subsection{Correlation function}
\label{subsec:result2}

We next consider the 
two-point correlation function, 
which can be computed from the power spectrum as
\begin{align}
\xi(r)=\int\frac{dk\,k^2}{2\pi^2}\,P(k)\,\frac{\sin(kr)}{kr}.
\label{eq:def_xi}
\end{align}
In Fig.~\ref{fig:xireal}, 
left panel focuses on the behaviors around the baryon acoustic 
peak, while right panel shows the global shape of the two-point 
correlation function plotted in logarithmic scales, for which $\xi(r)$
has been multiplied by the cube of the separation. The 
\RegPT~results agree with $N$-body simulations almost perfectly 
over the plotted scales. As it is known, the impact of non-linear 
clustering on the
baryon acoustic peak is significant:  
the peak position becomes slightly shifted to a smaller scale, and  
the structure of the peak tends to be smeared as 
the redshift decreases
(e.g., 
\cite{Eisenstein:2006nj,Crocce:2007dt,Smith:2007gi,Matsubara:2007wj}). 
The \RegPT~calculation can describe not only the behavior around 
the baryon acoustic peak but also the small-scale behavior of the correlation 
function. Note that similar results are also obtained from 
other improved PT treatments such as closure and LRT. 
Although the \RegPT~predictions eventually deviate 
from simulations at small scales -- the result at $z=0.35$ indeed 
manifests the discrepancy below $r\sim30\,h^{-1}$\,Mpc --
the actual range of agreement between \RegPT~and 
$N$-body results is even wider than what is naively 
expected from the power spectrum results. 
In fact, it has been recently advocated by several authors that with 
several improved PT treatments, 
the one-loop calculation is sufficient to 
accurately describe
the two-point correlation function 
(e.g., \cite{Okamura:2011nu,Taruya:2009ir,Reid:2011ar}). 
We have checked that the \RegPT~treatment at one-loop order can 
give a satisfactory result close to the two-loop result, and 
the prediction including the two-loop corrections only slightly improves 
the agreement with $N$-body simulations at small scales. 
This is good news for practical purposes in the sense that we do not 
necessarily have to evaluate the multi-dimensional 
integrals for the accurate prediction of two-point correlation function 
in the weakly non-linear regime. Nevertheless, in this work, we 
keep the two-loop contributions in the computed contributions.
The computational costs of the two-loop order will be addressed in the following 
with the development of a method for accelerated PT calculation at 
two-loop order.

\section{\RegPT{\tt -fast}: Accelerated power spectrum calculation}
\label{sec:PTreconst}

In this section, we present a method that allows accelerated calculations 
of the required diagrams of the two-loop order  \RegPT~prescription. 
In principle, the power spectra calculations in the context
of  \RegPT~require multi-dimensional integrations that cannot 
be done before-hand as they fully depend on the linear
power spectra. It is however 
possible to obtain the required quantities much more rapidly provided we 
know the answer for a close enough model.

The key point in this approach is to utilize the fact 
that the nonlinear \RegPT~power 
spectrum is a well-defined functional form of the linear power spectrum. Each of the diagram that has to be computed 
is of quadratic, cubic, etc. order with respect to the linear power 
spectrum with 
a kernel that, although complicated, can be explicitly given. It is then easy 
to Taylor expand each of these terms with respect to the linear power spectrum. 
In principle one then just needs  to prepare, in advance, 
a set of the \RegPT~results for some fiducial cosmological models, 
and then take the difference between fiducial and target initial power spectra 
for which 
we want to calculate the non-linear power spectrum. These differences involve
only one-dimensional integrals at the first order in the Taylor expansion.

In the following, we present the 
detail of the implementation of this approach 
illustrating it with the one-loop calculation case.

\subsection{Power spectrum reconstruction from fiducial model}
\label{subsec:PTreconst_1-loop}

While our final goal is to present the fast PT calculation at 
two-loop order, in order to get insights into the implementation of this 
calculation, we consider the power spectrum at 
one-loop order. The complete expressions needed for the fast PT calculation
at two-loop order, together with the prescription how to implement it,
is presented in Appendix \ref{sec:PTreconst_2-loop}.

Compared to the expressions given in 
Eq.~(\ref{eq:pk_Gamma_reg_2loop}), 
the power spectrum at one-loop order of the $\Gamma$-expansion 
reduces to
\begin{widetext}
\begin{align}
&P_{ab}(k;\eta) = \Gamma_{a,{\rm reg}}^{(1)}(k;\eta)
\Gamma_{b,{\rm reg}}^{(1)}(k;\eta)P_0(k)+
2\int\frac{d^3\bfq}{(2\pi)^3}\,
\Gamma_{a,{\rm reg}}^{(2)}(\bfq,\bfk-\bfq;\eta)
\Gamma_{b,{\rm reg}}^{(2)}(\bfq,\bfk-\bfq;\eta)P_0(q)P_0(|\bfk-\bfq|)
\label{eq:pk_Gamma_reg_1loop}
\end{align}
\end{widetext}
with the regularized propagators $\Gamma^{(1)}_{\rm reg}$ and 
$\Gamma^{(2)}_{\rm reg}$ valid at one-loop order being: 
\begin{align}
&\Gamma_{a,{\rm reg}}^{(1)}(k;\eta)=e^{\eta}\left[
1+\frac{k^2\sigmav^2e^{2\eta}}{2}+
e^{2\eta}\,\overline{\Gamma}^{(1)}_{a,{\rm 1\mbox{-}loop}}(k)\right]
\nonumber\\
&\qquad\qquad\qquad\qquad\qquad\qquad
\times\exp\left\{-\frac{k^2\sigmav^2e^{2\eta}}{2}\right\}, 
\label{eq:Gamma1_reg_1loop}
\\
& \Gamma_{a,{\rm reg}}^{(2)}(\bfq,\bfk-\bfq;\eta)=e^{2\eta}
F_{a,{\rm sym}}^{(2)}(\bfq,\bfk-\bfq)\,
\nonumber\\
&\qquad\qquad\qquad\qquad\qquad\qquad
\times\exp\left\{-\frac{k^2\sigmav^2e^{2\eta}}{2}\right\}.
\label{eq:Gamma2_reg_tree}
\end{align}
Note that the quantity $\overline{\Gamma}^{(1)}_{a,{\rm 1\mbox{-}loop}}$ 
is defined in Eq.~(\ref{eq:Gamma-p_nloop}), and explicitly given by
\begin{align}
\overline{\Gamma}^{(1)}_{a,{\rm 1\mbox{-}loop}}(k) = 3
\int\frac{d^3\bfq}{(2\pi)^3}\,F_{a,{\rm sym}}^{(3)}(\bfq,-\bfq,\bfk) P_0(q). 
\label{eq:Gamma1_1loop}
\end{align}
Thus, in Eq.~(\ref{eq:pk_Gamma_reg_1loop}), there apparently appear  
two contributions which involve multi-dimensional integrals; 
$\overline{\Gamma}^{(1)}_{a,{\rm 1\mbox{-}loop}}$ in the regularized propagator 
$\Gamma_{a,{\rm reg}}^{(1)}$, and 
the second term at the right-hand side. Although these contributions 
are known to be further reduced to one-, and two-dimensional integrals 
(e.g., Refs~\cite{Bernardeau:2011dp, Crocce:2005xz,Taruya:2007xy}), 
respectively, for the sake of this presentation we keep the expressions as in their original form.

As it has been mentioned earlier, the key idea of accelerated calculation is to prepare 
a set of \RegPT~results for fiducial cosmological models. Let us
denote the initial power spectrum for fiducial 
cosmology by $P_{0,{\rm fid}}(k)$. And we denote the initial spectrum for 
the target cosmological model, for which we want to calculate the non-linear 
power spectrum, by $P_{0,{\rm target}}(k)$. 
For the moment, we assume that the difference between those spectra
is small enough. Then, we may write
\begin{align}
P_{0,{\rm target}}(k)=P_{0,{\rm fid}}(k)+\delta P_0(k).
\label{eq:PT_initial_pk}
\end{align}
Hereafter, we focus on the power spectrum of density field, $P_{11}$, 
and drop the subscript. Substituting the above expression into 
Eqs.~(\ref{eq:pk_Gamma_reg_1loop})-(\ref{eq:Gamma2_reg_tree}), 
the non-linear power spectrum for the target model is symbolically 
written as 
\begin{align}
&P_{\rm target}(k;\eta) = P_{\rm un\mbox{-}pert}[k,\eta,\sigma_{\rm d,target}; 
P_{0,{\rm fid}}(k)] 
\nonumber\\
&\qquad\qquad\qquad\qquad
+ P_{\rm corr}[k,\eta, \sigma_{\rm d,target};\delta P_0(k)]. 
\label{eq:PTreconst}
\end{align}
Here, the first term at the right hand side is the 
un-perturbed part of the one-loop power spectrum, which is nothing but
the expression (\ref{eq:pk_Gamma_reg_1loop}) adopting 
the initial power spectrum for fiducial model, $P_{0,{\rm fid}}(k)$, but 
with the cosmological dependence of the time variable, 
given by $\eta=\ln D(z)$, being calculated from the target model. 
Also, the dispersion of displacement field, $\sigmav$, 
should be replaced with the one for the target model, i.e., 
$\sigma_{\rm d,target}=[\int dq P_{0,{\rm target}}(q)/(6\pi^2)]^{1/2}$.
In each term of Eq.~(\ref{eq:pk_Gamma_reg_1loop}), 
the scale and time dependence can be separately treated, and thus
the un-perturbed power spectrum, $P_{\rm un\mbox{-}perturbed}$, is 
evaluated algebraically by summing up each contribution, for which we 
use the precomputed data set in evaluating the scale-dependent function.

In Eq.~(\ref{eq:PTreconst}), the contribution $P_{\rm corr}$ includes the 
non-linear corrections originating from the differences of 
initial power spectra between fiducial and target cosmological models. 
To first order in $\delta P_0$, we have 
\begin{widetext}
\begin{align}
&P_{\rm corr}[k,\eta,\sigma_{\rm d,target};\delta P_0(k)] = 
2 \Gamma_{\rm reg}^{(1)}(k;\eta)\,
\delta \Gamma_{\rm reg}^{(1)}(k;\eta)\,P_{0,{\rm fid}}(k) +
\left[\Gamma_{\rm reg}^{(1)}(k;\eta)\right]^2\,\delta P_0(k)
\nonumber\\
&\qquad\qquad\qquad\qquad\qquad\qquad\qquad\qquad
+ 4\int\frac{d^3\bfq}{(2\pi)^3}\,
\left[\Gamma_{\rm reg}^{(2)}(\bfq,\bfk-\bfq;\eta)\right]^2P_{0,{\rm fid}}(|\bfk-\bfq|)
\delta P_0(q)
\label{eq:dpk_Gamma_reg_1loop}
\end{align}
\end{widetext}
In the above expression, The quantity $\sigmav$ appearing  
in the propagators $\Gamma_{\rm reg}^{(1)}$ and 
$\Gamma_{\rm reg}^{(2)}$ should be evaluated with the linear power spectrum for
the target cosmological model. The perturbed propagator 
$\delta \Gamma_{\rm reg}^{(1)}$ is expressed as
\begin{align}
&\delta \Gamma_{\rm reg}^{(1)}(k;\eta)=e^{3\eta}\,
\delta \overline{\Gamma}^{(1)}(k)\, 
e^{-k^2\sigma_{\rm d,target}^2e^{2\eta}/2}
\end{align}
where the kernel of integral in $\delta \overline{\Gamma}^{(1)}$ is the same
one as in Eq.~(\ref{eq:Gamma1_1loop}), but we may rewrite it with
\begin{align}
\delta\,\overline{\Gamma}^{(1)}(k)=\int \frac{dq\,q^2}{2\pi^2}\,
L_1^{(1)}(q,k)\,\delta P_0(q)
\label{eq:delta_Gamma1}
\end{align}
with the kernel $L_1^{(1)}$ given by  
\begin{align}
L_1^{(1)}(q,k) =3\,
\int\frac{d^2\bfOmg_q}{4\pi}\,F_{1,{\rm sym}}^{(3)}(\bfq,-\bfq,\bfk). 
\label{eq:def_L1}
\end{align}
Since the kernel $L_1^{(1)}$ only includes the PT kernel whose cosmological 
dependence is extremely weak, we can separately 
prepare the numerical data set 
for $L_1^{(1)}$ in advance \footnote{Indeed, the kernel $L_1^{(1)}$ 
is analytically known, and the explicit expression is given in, e.g., 
Refs.~\cite{Bernardeau:2011dp, Crocce:2005xz,Taruya:2007xy}.}. 
Then, we can use it to compute
$\delta\overline{\Gamma}^{(1)}$ for arbitrary $\delta P_0$, 
where the remaining integral to be evaluated is reduced to 
a one-dimensional integral.

Furthermore, the integral in the last term of 
Eq.~(\ref{eq:dpk_Gamma_reg_1loop}) is rewritten with
\begin{align}
&\int\frac{d^3\bfq}{(2\pi)^3} 
\left[\Gamma_{\rm reg}^{(2)}(\bfq,\bfk-\bfq;\eta)\right]^2
P_{0,{\rm fid}}(|\bfk-\bfq|)\,\delta P_0(q) 
\nonumber\\
&\quad=e^{-k^2\sigma_{\rm d,target}^2e^{2\eta}}\,e^{4\eta}\,
\int\frac{dq\,q^2}{2\pi^2}X^{(2)}(q,k)\,\delta P_0(q)
\end{align}
with the function $X^{(2)}$ being
\begin{align}
&X^{(2)}(q,k)=\frac{1}{2}\int_{-1}^1d\mu\,
\left[F^{(2)}_{\rm sym}(\bfq,\bfk-\bfq)\right]^2\,
\nonumber\\
&\quad\qquad\qquad\qquad\times P_{0,{\rm fid}}(\sqrt{k^2-2kq\mu+q^2}),
\end{align}
where the variable $\mu$ is the directional cosine defined by 
$\mu=(\bfk\cdot\bfq)/(kq)$. In deriving the above expression, 
we used the symmetric property of $\Gamma^{(2)}_{\rm reg}$, i.e., 
$\Gamma^{(2)}_{\rm reg}(\bfk_1,\bfk_2)=\Gamma^{(2)}_{\rm reg}(\bfk_2,\bfk_1)$. 
Since the quantity $X^{(2)}(q,k)$ can be computed in advance,  
all the integrals involving the power spectrum $\delta P_0$ 
are shown to be effectively reduced to one-dimensional integrals. 
In other words the only remaining task is to evaluate one-dimensional 
integrals, which can be done very efficiently. 

The practical implementation of this method makes use of another important property
of the kernel functions. They indeed have a very simple dependence on a global rescaling of the
power spectrum, $P_{0,{\rm fid}}\,\to\,c\,\,P_{0,{\rm fid}}$. It is then possible, without extra numerical computation, to choose the fiducial model among a continuous set of models. The model we choose, that is the normalization factor $c$ we take, is such that the difference 
$\delta P_0(k)$ is as small as possible in the wave-modes of interest. As we will see in Sec.~\ref{subsec:reliable}
it makes the use of this method very efficient.

Note that although the treatment depicted here does not give much impact 
on the computational cost of the one-loop calculation, we will explicitly 
show in Appendix \ref{sec:PTreconst_2-loop}  that at two-loop order the PT 
corrections involving multi-dimensional integrals can be similarly reduced 
to one-dimensional integrals. In the following, we denote \RegPTfast~the 
implementation of this approach at two-loop order.

\begin{figure*}[t]


\includegraphics[width=8cm,angle=0]{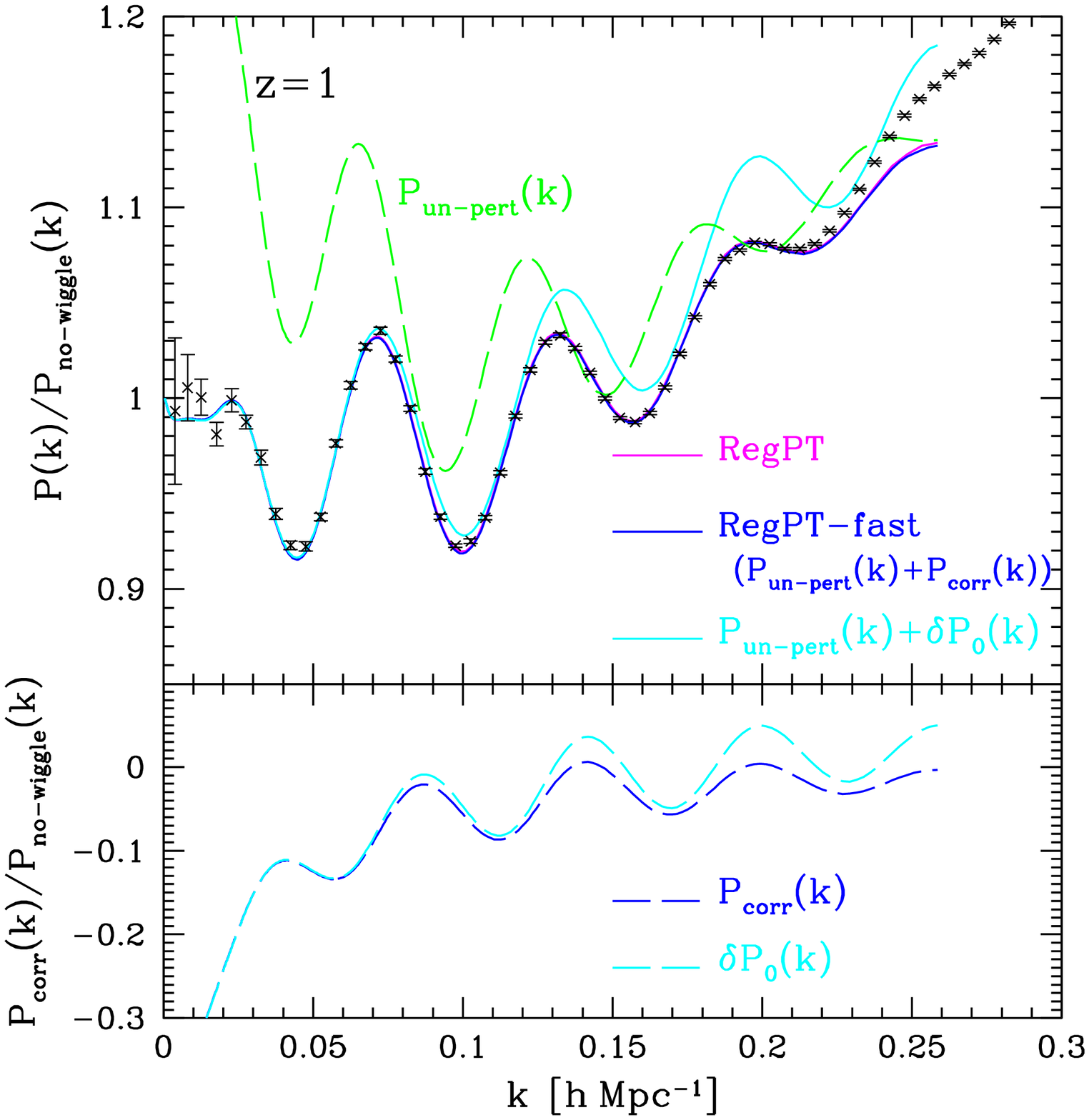}
\hspace*{0.0cm}
\includegraphics[width=8cm,angle=0]{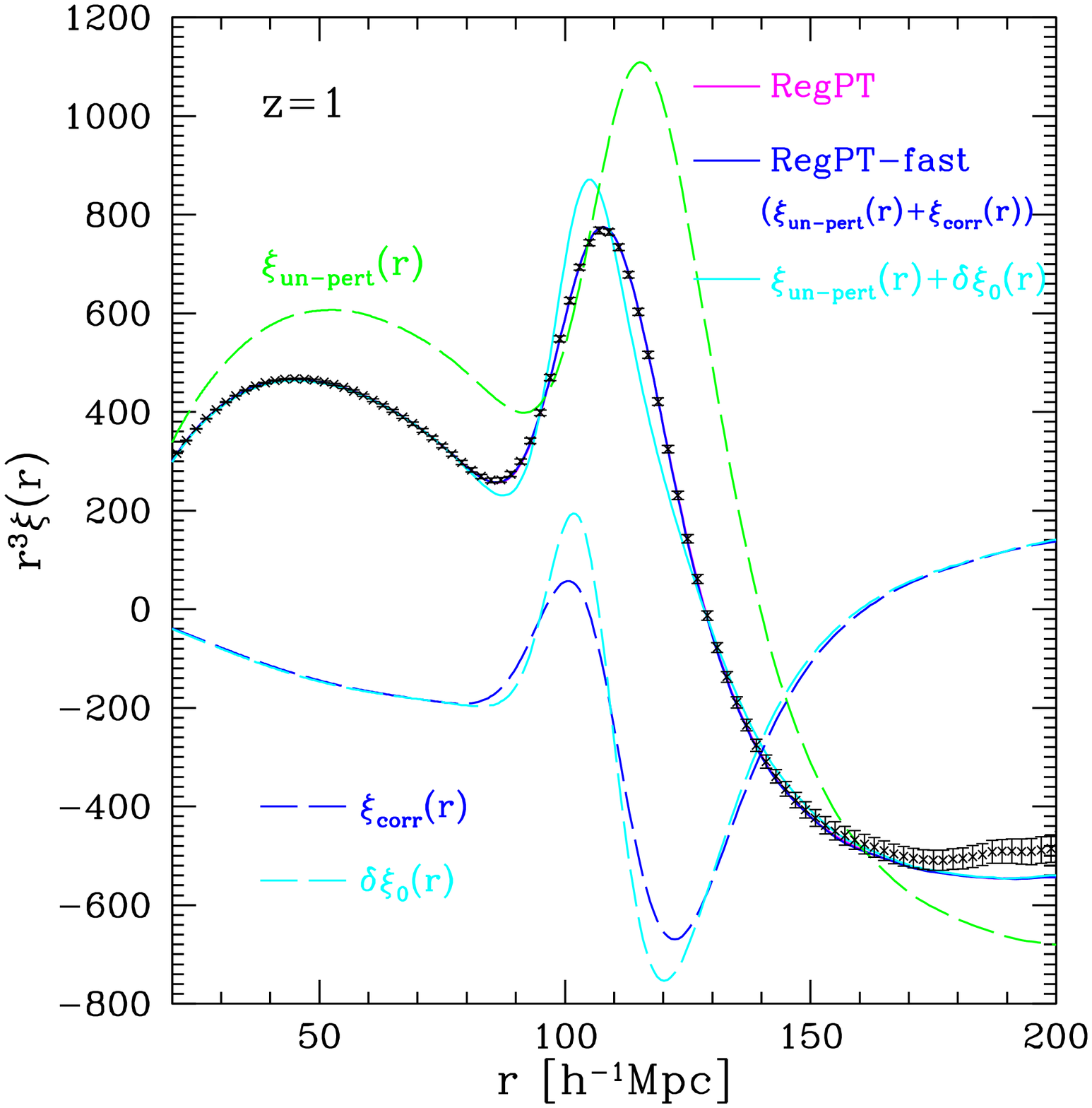}

\vspace*{0.5cm}

\caption{Example of the performances of the \RegPTfast~approach compared to direct  \RegPT~calculation. Left panel shows 
the power spectrum divided by the smooth reference spectrum, $P(k)/P_{\rm no\mbox{-}wiggle}(k)$, while right panel plots the correlation function multiplied by 
the cube of separation, i.e., $r^3\,\xi(r)$. In both panels, the 
results at $z=1$ are shown, together with $N$-body simulations. 
The \RegPTfast~results, 
computed with prepared data set for fiducial cosmological model,  
are plotted as magenta solid lines, which   
almost coincide with those obtained from the rigorous \RegPT~calculation (magenta solid). As shown in Eq.~\ref{eq:PTreconst}, 
the \RegPTfast~results are divided into two contributions; 
un-perturbed part ($P_{\rm un\mbox{-}pert}$ or $\xi_{\rm un\mbox{-}pert}$) 
adopting the {\tt wmap3} model as fiducial cosmology,  
and the correction part ($P_{\rm corr}$ or $\xi_{\rm corr}$) 
evaluated with the power spectrum difference $\delta P_0$. 
These are respectively plotted as green dashed and cyan dashed lines.  
For reference, we also show the linearly evolved result of 
power spectrum difference $\delta P_0$ (cyan dashed) 
and the sum of the contributions $P_{\rm un\mbox{-}pert}+\delta P_0$ (cyan solid) 
in left panel, and their Fourier counterparts in right panel. 
\label{fig:pk_xi_RegPTfast_zred1}}
\end{figure*}

\subsection{Performances}
\label{subsec:demonstration}

Let us now illustrate the efficiency of the \RegPTfast~expansion. Based 
on the expressions given in Appendix~\ref{sec:PTreconst_2-loop},  we 
calculate the power spectrum and correlation function  at two-loop order. 
We adopt the best-fit parameters determined by the  third-year WMAP 
result \cite{Spergel:2006hy} 
as the fiducial cosmological model from which we try to reproduce 
the \RegPT~results for the five-year WMAP 
cosmological model. Cosmological parameters for the fiducial model is 
listed in Table \ref{tab:fiducial_model}. Compared to the target model 
in Table \ref{tab:nbody_nishimichi}, the mass density  parameter shows 
a $20\%$ difference, and with $7\%$ enhancement in the 
power spectrum normalization ($\sigma_8$),  this leads to a $20$-$30$\% 
difference in the initial power spectrum.

Fig.~\ref{fig:pk_xi_RegPTfast_zred1} plots the results of
the \RegPTfast~calculation (blue) compared to the target \RegPT~calculation (magenta). We plot, for a specific redshift $z=1$,
the ratio of the power spectrum to the smooth
reference spectrum, $P(k)/P_{\rm no\mbox{-}wiggle}(k)$, 
and correlation function multiplied by the 
cube of separation, $r^3\xi(r)$, in left and right panels, respectively. 
The \RegPTfast~results perfectly coincide with  \RegPT~direct calculation, even outside the range of agreement with $N$-body 
simulations.

Note that the perfect match between \RegPT~and \RegPTfast~results 
is due to a large extent to the contributions of 
the higher-order PT in the correction, $P_{\rm corr}$ or $\xi_{\rm corr}$.  
This appears clearly in the plots of the linear theory correction, 
$\delta P_0=P_{0,{\rm target}}-P_{0,{\rm fid}}$ and its Fourier counterpart 
$\delta\xi_0$ (cyan long-dashed). As shown in cyan solid lines,   
the total contribution, i.e., the combination of the un-perturbed part 
plus linear theory correction, somehow resembles the result with 
direct \RegPT~calculation, but exhibits a rather prominent oscillatory feature 
with slightly different phase in power spectrum, leading 
to a non-negligible discrepancy. Accordingly, in correlation function, 
the acoustic peak becomes enhanced, and the position of peak is shifted to 
a small separation. 
Note  finally, that these results could only be achieved with the help 
of the rescaling properties 
of the kernel functions. In this particular case the fiducial model has 
been rescaled as
$P_{0,{\rm fid}}\to 1.082\,P_{0,{\rm fid}}$. Rescaling is a key feature of the 
\RegPTfast~method. It will be further discussed in the next section.

\begin{table*}[t]
\caption{\label{tab:fiducial_model} Cosmological parameters for fiducial 
models used for the RegPT-fast calculation 
(see Sec.~\ref{subsubsec:convergence}).} 
\begin{ruledtabular}
\begin{tabular}{lccccccc}
Name & 
$\Omega_{\rm m}$ & $\Omega_{\Lambda}$ & $\Omega_{\rm b}/\Omega_{\rm m}$ & $w$ & $h$ & $n_s$ & $\sigma_8$ \\
\hline
\verb|wmap3| & 0.234 &  0.766 & 0.175 & -1 & 0.734 & 0.961 & 0.760\\
\verb|M001| & $0.430_7$ &  $0.569_2$ & 0.150 & -0.816 & $0.597_7$ & $0.946_8$ & $0.816_1$ \\
\verb|M023| & $0.160_2$ &  $0.839_8$ & $0.181_7$ & -1.261 & $0.869_4$ & $0.901_6$ & $0.666_4$ \\
\end{tabular}
\end{ruledtabular}
\end{table*}

\begin{figure}[b]


\includegraphics[width=10cm,angle=0]{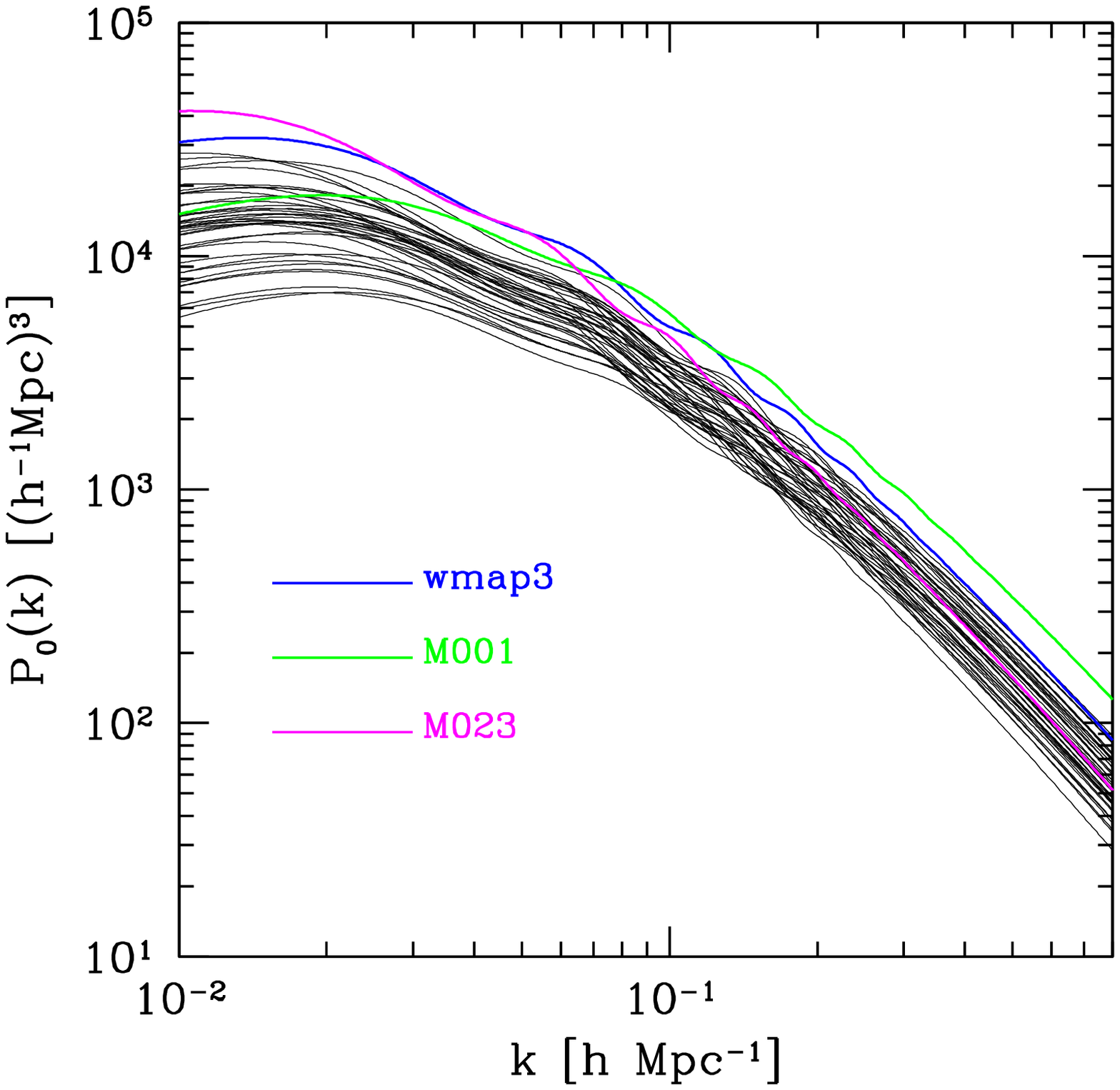}

\vspace*{-1.5cm}

\caption{Linear power spectra $P_0(k)$ for 38 cosmological 
  models \cite{Lawrence:2009uk}. \Blue{Blue}, \Green{green}, and 
  \Magenta{magenta} lines are respectively 
  the power spectra of fiducial models {\tt wmap3}, {\tt M001}, 
  and {\tt M023} used for the \RegPTfast calculation 
  (see Table \ref{tab:fiducial_model} for their cosmological parameters). 
\label{fig:pklin_coyote}}
\end{figure}

\section{Testing \RegPT~treatment for varying cosmological models}
\label{subsec:reliable}

The purpose of this section is twofold.  
Our first goal is to explore the validity and applicability of the  
\RegPTfast~scheme. 
Having shown that the \RegPTfast~approach can be used in one
specific example,  we now want to discuss the usefulness of this treatment 
from a more practical 
point of view. To be precise, we want to know how well the 
\RegPTfast~treatment can reproduce 
rigorous \RegPT~calculation in a variety of cosmological models. 

Our second and natural goal is to test the \RegPT~scheme itself, 
whether from direct or fast calculations, against 
N-body based predictions such that the {\tt cosmic emulator}\footnote{{\tt 
http://www.lanl.gov/projects/cosmology/CosmicEmu/}}.

To do that, we have selected the $38$ cosmological models investigated in 
Ref.~\cite{Lawrence:2009uk} for which we can use the publicly released code, 
{\tt cosmic emulator}, that provides interpolated power spectra derived 
from $N$-body simulations. Let us remind that the cosmological models 
considered there are sampled from a wide parameter 
space for flat $w$-CDM cosmology, and lie within the range,
\begin{align}
0.120<\Omega_{\rm m}h^2<0.155,
\nonumber\\
0.0215<\Omega_{\rm b}h^2<0.0235,
\nonumber\\
0.85<n_{\rm s}<1.05,
\nonumber\\
-1.30<w<-0.70,
\nonumber\\
0.616<\sigma_8<0.9.  \nonumber 
\end{align}
The concrete values of the cosmological parameters in each model are not 
shown here. Readers can find them 
in Table 1 of Ref.~\cite{Lawrence:2009uk}. 
Fig.~\ref{fig:pklin_coyote} shows the linear power spectra $P_0(k)$ for 
the $38$ cosmological models, which have been all 
produced with CMB Boltzmann code,  {\tt camb}~\cite{Lewis:1999bs}.

\subsection{Convergence of \RegPTfast}
\label{subsubsec:convergence}

\begin{figure*}[tb]

\vspace*{-2cm}

\includegraphics[width=12cm,angle=0]{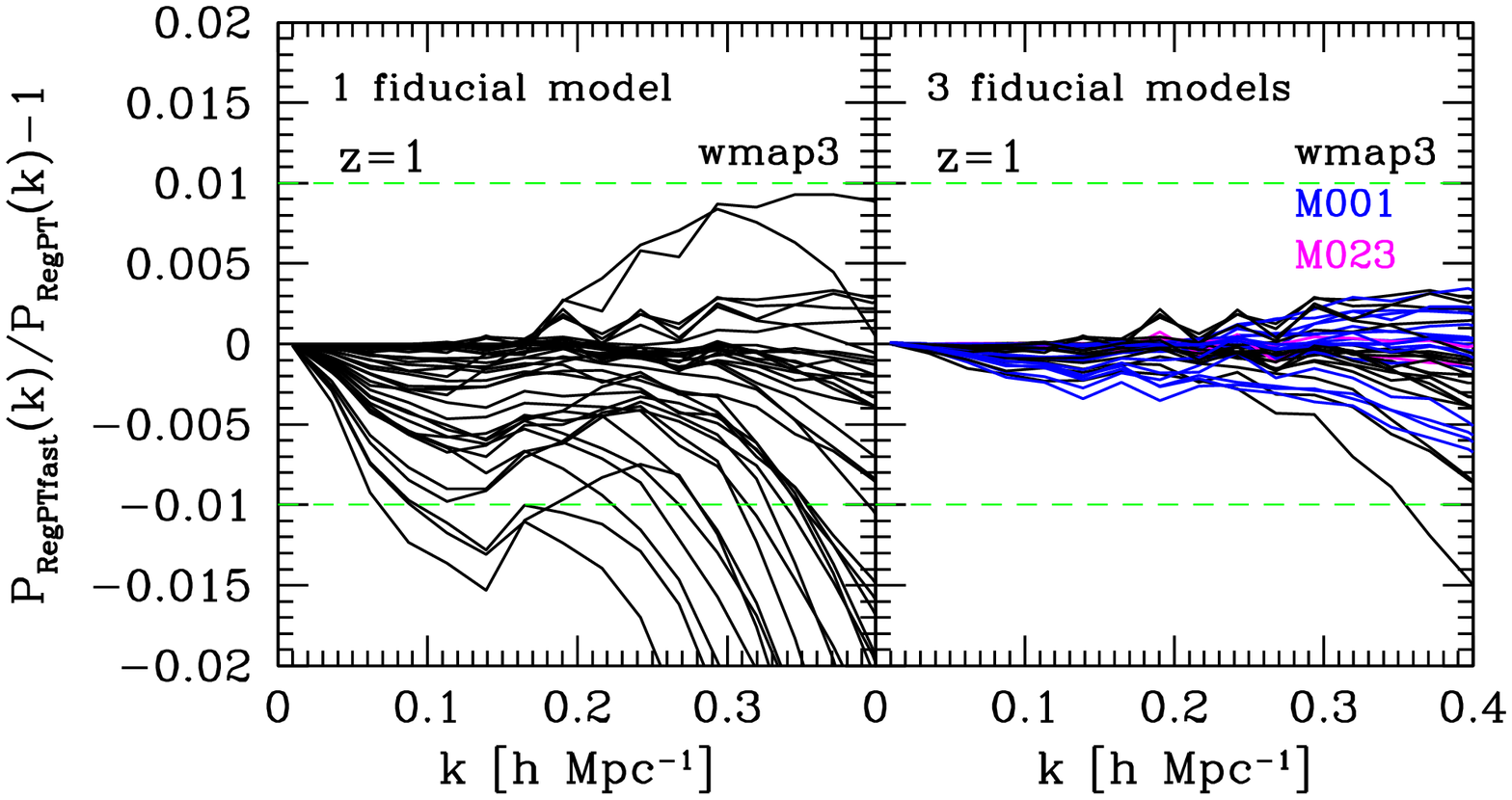}

\vspace*{-3cm}

\caption{Convergence between the \RegPTfast~treatment and the direct \RegPT~calculations for 
$38$ cosmological models. The fractional difference of the power spectra 
between \RegPTfast~and \RegPT~direct calculations, 
$P_{\rm RegPTfast}(k)/P_{\rm RegPT}(k)-1$, is plotted at 
$z=1$. Left and right panels respectively show the results adopting the 
one and three fiducial models. 
\label{fig:check_coyote}}
\end{figure*}
\begin{figure*}[ht]

\vspace*{-0.5cm}

\includegraphics[width=16cm,angle=0]{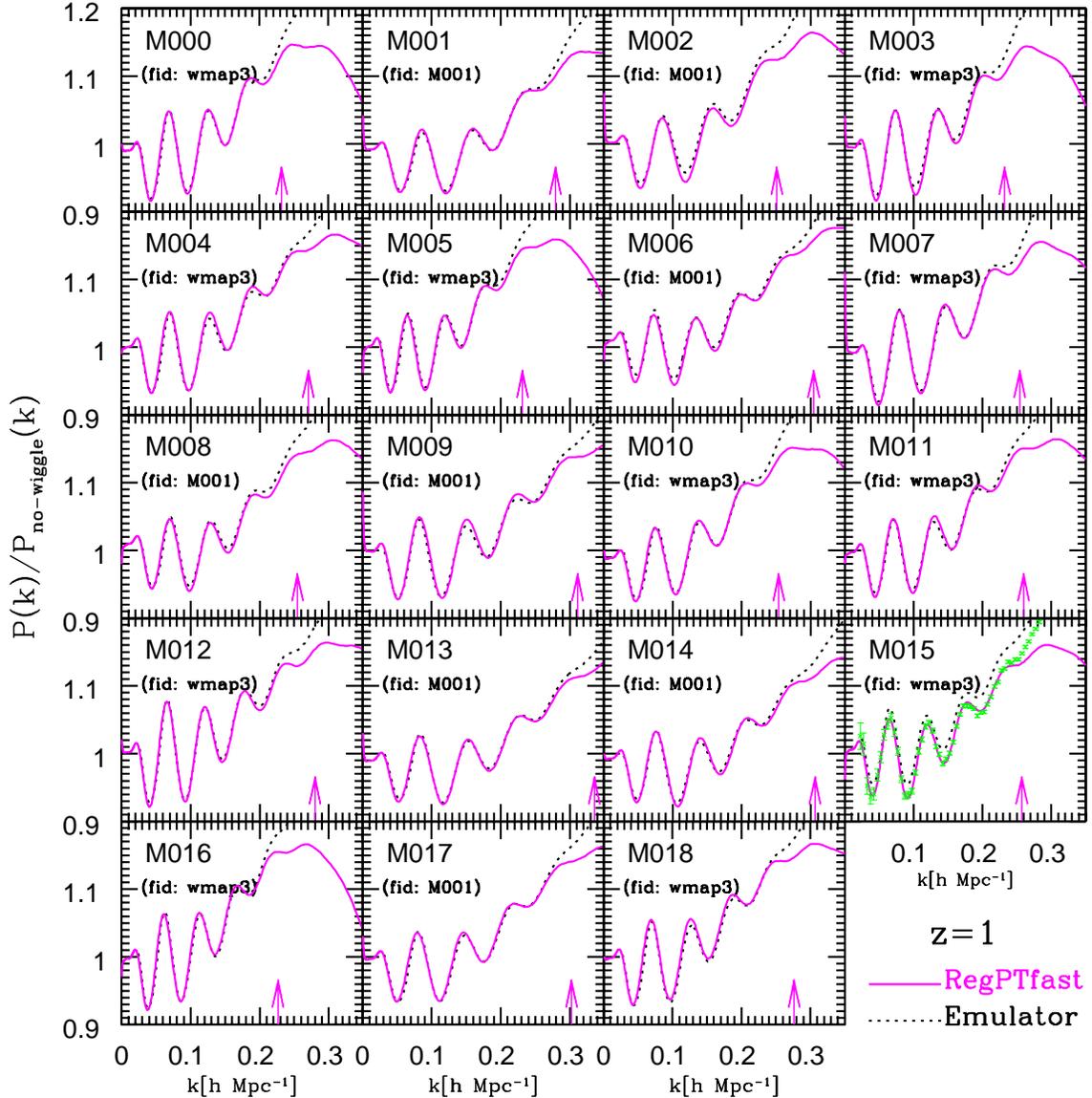}

\vspace*{-0.5cm}

\caption{Ratio of power spectra, $P(k)/P_{\rm no\mbox{-}wiggle}(k)$, at $z=1$ 
  for the cosmological models M000--M017. Solid and dotted lines are 
  obtained from the \RegPTfast~and cosmic emulator codes, respectively. The 
  fiducial model used for the \RegPTfast~calculation is indicated in each 
  panel.  The vertical arrows mean the critical wavenumber $k_{\rm crit}$ 
  defined by Eq.~(\ref{eq:k_crit}), which roughly gives 
  an applicable range of \RegPT~prediction (see text). 
\label{fig:ratio_coyote}}
\end{figure*}
\begin{figure*}[ht]

\vspace*{-0.5cm}

\includegraphics[width=16cm,angle=0]{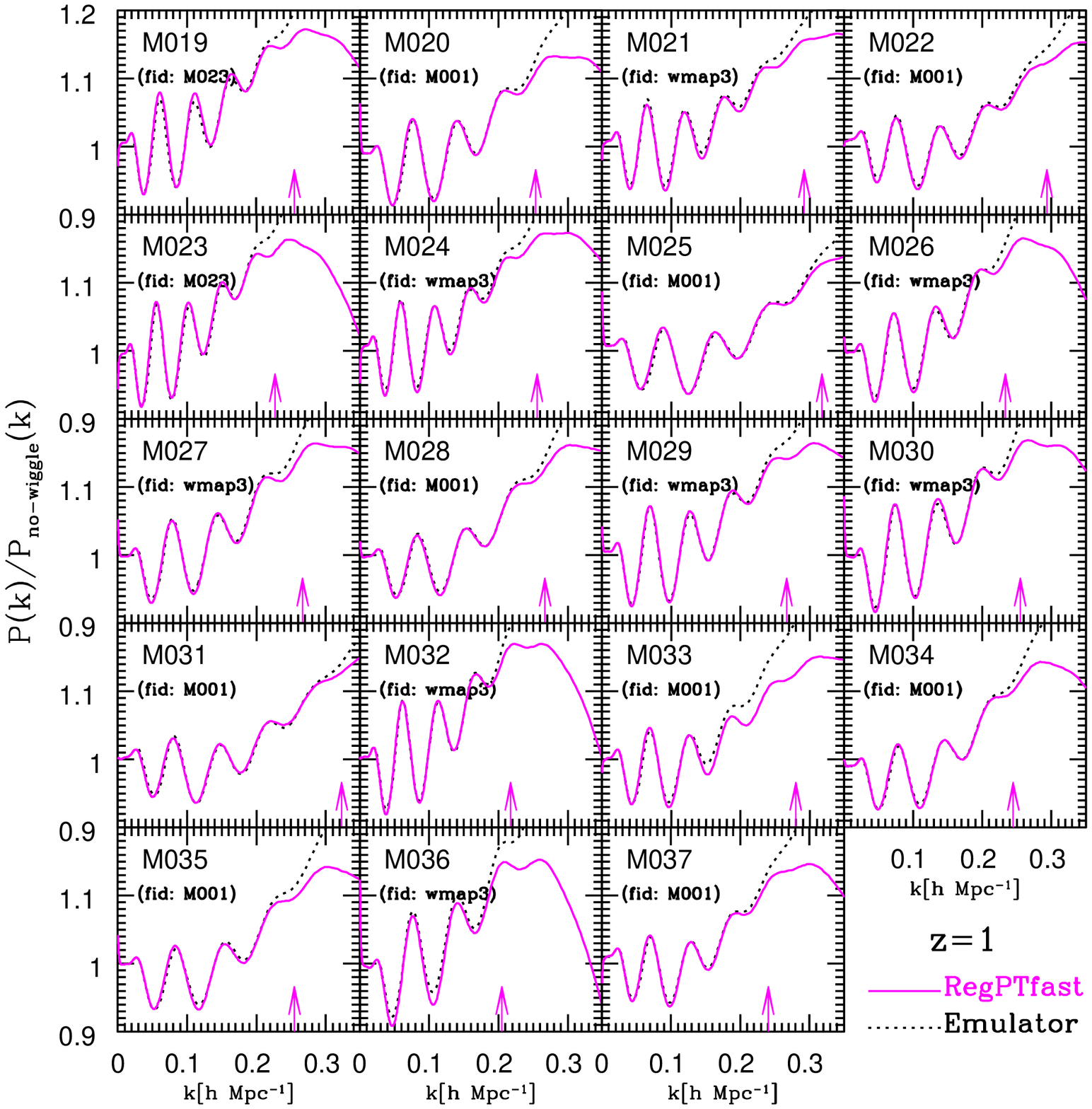}

\vspace*{-0.5cm}

\caption{Same as Fig.~\ref{fig:ratio_coyote}, but for the models 
M019--M037. 
\label{fig:ratio_coyote2}}
\end{figure*}

Let us first examine the convergence of the power spectrum calculations 
between \RegPT~and \RegPTfast~treatments. 
We ran both the \RegPTfast~and \RegPT~codes, and 
evaluate the fractional difference between 
these power spectra, defined by $P_{\rm RegPTfast}(k)/P_{\rm RegPT}(k)-1$. 
Collecting the results at $z=1$ in each cosmological model, 
the convergence of the power spectrum calculations for $38$ models 
is summarized in Fig.~\ref{fig:check_coyote}. 
In left panel we show the result when only one fiducial model, {\tt wmap3}, 
is used. In that case \RegPTfast~results tend to 
underestimate the results from rigorous \RegPT~calculations at increasing $k$, 
and most of them eventually exceed the $1\%$ difference, indicated by 
the green dashed line. This is 
because the shape of the initial power spectrum in each target model 
is rather different from that in the fiducial model, and even 
adjusting the re-scaling parameter $c$ cannot compensate a large 
power spectrum difference.
To be more precise, in most of the models, 
the shape parameter, defined by $\Gamma=\Omega_{\rm m}h$, 
is typically larger than 
the one in the fiducial model. As a consequence, even if we adjust the 
power spectrum at large scales to match the one in  
the target model, 
the difference $|\delta P_0|$ can become large as increasing $k$,  
leading to a failure of the perturbative reconstruction by \RegPTfast. 

To remedy this situation, a simple but efficient approach is to 
enlarge our set of fiducial models with $\Gamma$-parameters 
that differ from the one of  {\tt wmap3}  model, i.e. $\Gamma=0.172$.
Right panel of  Fig.~\ref{fig:check_coyote} shows the convergence results when we 
supply two extra fiducial models whose cosmological 
parameters are listed in Table \ref{tab:fiducial_model}. 
As a fiducial model with a larger shape parameter,  
we adopt the {\tt M001} cosmological model ($\Gamma=0.257$). 
Further, for a
secure calculation applicable to general cosmological models, 
we also supply another fiducial model, 
{\tt M023}, which has a smaller shape parameter ($\Gamma=0.139$). 
The initial power spectra of those models are plotted in 
Fig.~\ref{fig:pklin_coyote}, depicted as 
green ({\tt M001}) and magenta ({\tt M023}) solid lines.  
As a result, the convergence of the 
power spectrum calculations is dramatically 
improved, and the \RegPTfast~now coincides with rigorous \RegPT~calculation 
with $\lesssim0.4\%$ precision at $k\lesssim0.3\,h$\,Mpc$^{-1}$. 
Although there still exist exceptional cases, in which the fractional 
difference eventually exceeds $1\%$ precision at 
$k\gtrsim0.36\,h$\,Mpc$^{-1}$, in practice this is beyond the applicable range
of the \RegPT~calculation itself. 

With this setting, making use of these three fiducial models, \RegPTfast~reproduces \RegPT~direct
calculations in a wide range of cosmological models and, also it does not appear here, for a redshift range
of general interest.

\subsection{Comparison with cosmic emulator}
\label{subsubsec:emulator}

It is now time to discuss the accuracy of the overall
\RegPT~scheme  with general {\tt cosmic emulator} predictions. 
Figs.~\ref{fig:ratio_coyote} and \ref{fig:ratio_coyote2} summarize
the results of the comparison for all $38$ models, where  
we plot the ratios of power spectra, $P(k)/P_{\rm no\mbox{-}wiggle}(k)$, 
at specific redshift $z=1$. In each panel, 
magenta solid and black dashed lines represent the results 
of \RegPTfast~and the power spectrum emulator code, 
respectively. 
Also, the fiducial model used for the \RegPTfast~calculation is 
indicated, together with the label of the cosmological model. 
The two results mostly coincide with each other, and are hardly 
distinguishable at $k\lesssim0.2\,h$\,Mpc$^{-1}$, where 
the linear theory prediction typically produces a $10\%$ error. 
At $k\gtrsim0.2\,h$\,Mpc$^{-1}$, the \RegPTfast~results tend to 
deviate from the predictions of the emulator code which probably indicates the
the limitation of PT treatment.  However, some models still show a remarkable agreement at $k\lesssim0.3\,h$\,Mpc$^{-1}$ (e.g., M009 and M013).

As the range of applicability of the \RegPT~scheme depend on both $k$ and the power spectrum
amplitude, following Refs.  \cite{Nishimichi:2008ry,Taruya:2009ir},  
we propose here a phenomenological rule for the domain of applicability of the \RegPT~
calculations. The proposed upper value for $k$ is $k_{\rm crit}$ that can be obtained
from the implicit equation,
\begin{align}
\frac{k_{\rm crit}^2}{6\pi^2}\int_0^{k_{\rm crit}} dq\,P_{\rm lin}(q;z)=C,
\label{eq:k_crit}
\end{align}
where $C$ is a fixed constant, $C=0.7$. The resulting 
values are
depicted as vertical arrows in Figs.~\ref{fig:ratio_coyote} and 
\ref{fig:ratio_coyote2}. Below the critical wavenumber, the
\RegPT~scheme indeed agrees with results of the
emulator code, 
mostly within a percent-level precision\footnote{We however noticed that some models exhibit 
non-negligible discrepancy between the results of 
\RegPTfast~and the emulator codes, even well below $k_{\rm crit}$.   
One of such is M015, 
showing a broad-band discrepancy over the plotted range. 
This is somewhat surprising in the sense that the \RegPTfast~result
almost converges the linear theory prediction 
at $k\lesssim0.12\,h$\,Mpc$^{-1}$, while the result of the
emulator code is still away from it. 
To better understand the source of 
the discrepancy,  
we have ran $N$-body simulations for the M015 model -- cosmological 
parameters of M015 model were set as 
$\Omega_{\rm m}=0.2364$, $\Omega_{\rm b}=0.0384$, $w=-1.281$, $h=0.7737$, 
$n_s=1.0177$, and $\sigma_8=0.7692$. --  
with the same setup as listed in Table~\ref{tab:nbody_nishimichi}. 
The resulting power spectrum, estimated from the ensemble of 
the $8$ independent realizations, is superposed in the panel of M015 
(green symbols with errorbars) and is shown 
to faithfully trace the \RegPTfast~result up to the critical wavenumber. 
It points to a possible flaw in the power spectrum emulator, 
in estimating the smooth power spectrum from  
the ensemble of simulation results, or constructing the interpolated 
result of the simulated power spectra.}. We have also checked that this is also the case for $z=0.5$
with this definition of $k_{\rm crit.}$.

The  \RegPT~scheme is therefore shown to give a fairly accurate prediction for 
the power spectrum in the weakly non-linear regime in the sense given above. 
\RegPT~direct calculations, or (almost) 
equivalently,  
\RegPTfast~calculations with the three fiducial models we prepared,
can be applied to a wide range of cosmological models.
Though we did not discuss 
it here, we expect the same to be also true for the correlation function. 
Finally we note that as the relevant scale of weakly non-linear regime grows wider for higher redshifts, the applicability and reliability 
of the \RegPT~scheme is naturally enhanced.  On the other hand, the emulation schemes to build up interpolated results
from large sets of $N$-body simulations are generally efficient 
in predicting the power spectrum at non-linear scales but are more likely to fail at high-$z$, since 
the requirement for the force resolution in $N$-body simulation 
becomes more and more severe. 
In this respect, perturbative reconstruction schemes such as \RegPT~-- but this would also be the case
of \mptbreeze~--  are complementary to $N$-body based predictions.

\section{Conclusion}
\label{sec:conclusion}

It is needless to say that 
future cosmological observations make 
the development of cosmological tool aiming at accurately 
predicting the large-scale statistical properties of the universe highly desirable. 
In the first part  of the present paper, based on a
renormalized perturbation 
theory (PT), we introduced an explicit computation scheme applied to 
the matter power spectrum and correlation function 
in weakly non-linear regime that consistently includes the 
PT corrections up to the two-loop order. 
The construction of the full expression for the power spectrum is based on
the $\Gamma$-expansion, i.e.  makes use of the multi-point propagators 
which are properly regularized so as to recover 
their expected resummed behavior at high-$k$ and to match the standard 
PT result at low-$k$. We call this regularized PT treatment \RegPT. We 
have shown that the \RegPT~scheme provides 
an accurate prediction for both the power spectrum and 
the correlation function, 
leading to a percent-level agreement with $N$-body simulations in the 
weakly non-linear regime.

In the second half of the paper, we presented a method to 
accelerate the power spectrum calculations. 
The method utilizes prepared data sets for some specific fiducial models from which regularized PT 
calculations can be performed for arbitrary cosmological models.
The main interest of this method is that the evaluation 
the residual PT corrections between fiducial and target cosmological 
models can be reduced to mere one-dimensional  integrals. 
This enables us to dramatically  reduce the computational 
cost, and even with single-node calculation by a
laptop computer, the
power spectrum calculation can be done in a few seconds. 
We call this method \RegPTfast, and we have demonstrated that the \RegPTfast~treatment can 
perfectly reproduce the direct \RegPT~calculations that involve several multi-dimensional integrals.

We then investigated the range of applicability of the \RegPT~schemes in a broad 
class of cosmological models.
For this purpose, we select 38 cosmological models, and compared 
the \RegPT~predictions -- eventually incorporating the accelerated computations -- with results 
of a 
power spectrum emulator code, {\tt cosmic emulator}. We show that with the
help of  three fiducial models the \RegPTfast~calculations give reliable predictions for the power spectra 
over this range of cosmological models\footnote{Our analysis is however restricted to flat $w$-CDM 
models.}. We furthermore put forward an empirical criterion 
(\ref{eq:k_crit}) that gives a good indication of the
applicable range of the 
\RegPT~scheme in $k$. The \RegPTfast~treatment, 
together with the
direct \RegPT~calculation, has been implemented 
in a fortran code that we publicly release as part 
of this paper.

Although this paper is focused on precision calculations of the 
matter power spectrum, the \RegPT~framework as well as the methodology for accelerated 
calculation can naturally be applied to the power spectrum of the velocity divergence 
and the cross-power spectrum of velocity and density fields in a similar way. 
The analysis of the velocity power spectrum, 
together with a detailed comparison with $N$-body simulations, will be 
presented elsewhere. Of particular interest 
is the application of the \RegPT~schemes  to the 
redshift-space power spectrum or correlation function. In this case, 
not only the velocity and density power spectra, 
but also the multi-point spectra like bispectrum, 
arising from the non-linear mode coupling, seem to play 
important roles, and should be properly modeled. Significance of the effect 
of multi-point spectra has been recently advocated by 
Refs.~\cite{Reid:2011ar,Taruya:2010mx,Nishimichi:2011jm,Tang:2011qj}, 
and there appear physical models that account for this. 
Combination of these models with 
the \RegPT~schemes would be very
important, and we will discuss it in a near future.

\begin{acknowledgments}
A.T. is grateful to Toshiya Namikawa for his helpful comments 
on the code implentation. 
S.C. thanks Christophe Pichon for fruitful comments and discussion. 
This work has been benefited from exchange visits supported by a 
bilateral grant from 
Minist\`ere  Affaires Etrang\`eres et Europ\'eennes in France
and Japan Society for the Promotion of Science (JSPS).  
A.T. acknowledges support from Institutional Program for Young
Researcher Overseas Visit funded by the JSPS. 
A.T. is also supported in part by a Grant-in-Aid for Scientific 
Research from the JSPS (No.~24540257). 
T. N. is supported by a Grant-in-Aid for JSPS Fellows (PD: 22-181) and 
by World Premier International Research Center Initiative
(WPI Initiative), MEXT, Japan. Numerical computations for the present work 
have been
carried out in part on Cray XT4 at Center for Computational Astrophysics, 
CfCA, of National
Astronomical Observatory of Japan, and in part under the Interdisciplinary 
Computational 
Science Program in Center for Computational Sciences, University of Tsukuba.
F.B and S.C. are also partly supported by the French Programme National 
de Cosmologie et Galaxies. 
\end{acknowledgments}

\appendix

\section{Perturbative reconstruction of \RegPT~ power spectrum at 
two-loop order}
\label{sec:PTreconst_2-loop}

In this Appendix, we present the set of perturbative expressions
that are used for the accelerated 
power spectrum calculation at two-loop order
which is implemented in the \RegPTfast~code.

In a similar manner to the one-loop case described in 
Sec.~\ref{subsec:PTreconst_1-loop}, we can expand the power spectrum
expression up to two-loop order around the fiducial cosmological model, 
and obtain the perturbative expression for power spectrum in the target 
cosmological model. 
Plugging Eq.~(\ref{eq:PT_initial_pk}) into the two-loop expression  
 (\ref{eq:pk_Gamma_reg_2loop}) and assuming $\delta P_0\ll P_{0,{\rm fid}}$, 
 the power spectrum is written like (\ref{eq:PTreconst}), and 
 the correction $P_{\rm corr}$ becomes
\begin{widetext}
\begin{align}
&P_{\rm corr}[k,\eta,\sigma_{\rm d,target};\delta P_0(k)] = 
2 \Gamma_{\rm reg}^{(1)}(k;\eta)\,
\delta \Gamma_{\rm reg}^{(1)}(k;\eta)\,P_0(k) +
\left[\Gamma_{\rm reg}^{(1)}(k;\eta)\right]^2\,\delta P_0(k) 
\nonumber\\
&\qquad
+4\int\frac{d^3\bfq}{(2\pi)^3}\,
\left\{
\left[\Gamma_{\rm reg}^{(2)}(\bfq,\bfk-\bfq;\eta)\right]^2P_0(|\bfk-\bfq|)\,
\delta P_0(q) +
\Gamma_{\rm reg}^{(2)}(\bfq,\bfk-\bfq;\eta)\,
\delta\Gamma_{\rm reg}^{(2)}(\bfq,\bfk-\bfq;\eta)
\,P_0(|\bfk-\bfq|)\,P_0(q)
\right\}
\nonumber\\
&\qquad
+18\int\frac{d^3\bfp d^3\bfq}{(2\pi)^6}\,
\left[\Gamma_{\rm reg}^{(3)}(\bfp,\bfq,\bfk-\bfp-\bfq;\eta)\right]^2
P_0(|\bfk-\bfp-\bfq|)\,P_0(p)\,
\delta P_0(q). 
\label{eq:dpk_Gamma_reg_2loop}
\end{align}
\end{widetext}
In the above, the perturbations of regularized propagators, 
$\delta\Gamma_{\rm reg}^{(1)}$ and $\delta\Gamma_{\rm reg}^{(2)}$, are
described as 
\begin{align}
&\delta\Gamma_{\rm reg}^{(1)}(k;\eta)=
e^{3\eta}\left[
(1+\alpha_k)\,
\delta\overline{\Gamma}^{(1)}_{\rm 1\mbox{-}loop}(k) +
e^{2\eta}\,
\delta\overline{\Gamma}^{(1)}_{\rm 2\mbox{-}loop}(k)\right]
\nonumber\\
&\qquad\qquad\quad
\times e^{-\alpha_k},
\\
&\delta\Gamma_{\rm reg}^{(2)}(\bfq,\bfk-\bfq;\eta)=
e^{4\eta}
\,\delta\overline{\Gamma}^{(2)}_{\rm 1\mbox{-}loop}(\bfq,\bfk-\bfq)
\,e^{-\alpha_k},
\end{align}
where we define $\alpha_k\equiv k^2\sigma_{\rm d,target}^2e^{2\eta}/2$. 
The quantities $\delta\overline{\Gamma}^{(p)}_{\rm n\mbox{-}loop}$ 
are defined by
\begin{align}
&\delta\overline{\Gamma}^{(1)}_{\rm 1\mbox{-}loop}(k) =
3\,\int\frac{d^3\bfq}{(2\pi)^3}\,
F_{\rm sym}^{(3)}(\bfq,-\bfq,\bfk)\,\delta P_0(q),
\\
&\delta\overline{\Gamma}^{(1)}_{\rm 2\mbox{-}loop}(k) =
30\,\int\frac{d^3\bfq_1 d^3\bfq_2}{(2\pi)^6}\,
F_{\rm sym}^{(5)}(\bfq_1,-\bfq_1,\bfq_2,-\bfq_2,\bfk)
\nonumber\\
&\qquad\qquad\qquad\quad\times P_{0,{\rm fid}}(q_1)\,\delta P_0(q_2),
\\
&\delta\overline{\Gamma}^{(2)}_{\rm 1\mbox{-}loop}(\bfk_1,\bfk_2) =6
\int\frac{d^3\bfq}{(2\pi)^3}\,F_{\rm sym}^{(4)}(\bfq,-\bfq,\bfk_1,\bfk_2)\,
\delta P_0(q).
\end{align}
The kernels $F_{\rm sym}^{(p)}$ are the symmetrized standard PT kernel for
density field. In the above, the angular integrals are known to be 
analytically performed 
(Refs.~\cite{Crocce:2005xz,Bernardeau:2011dp}, and Bernardeau et al. in prep.), 
one may write 
\begin{align}
&\delta\overline{\Gamma}^{(1)}_{\rm 1\mbox{-}loop}(k) =
\int\frac{dq\,q^2}{2\pi^2}\,f(q;k)\,\delta P_0(q),
\\
&\delta\overline{\Gamma}^{(1)}_{\rm 2\mbox{-}loop}(k) =
2\,\int\frac{dq_1 \,dq_2\,\,q_1^2\,q_2^2}{(2\pi^2)^2}\,J(q_1,q_2;k)
\nonumber\\
&\qquad\qquad\qquad\qquad\qquad
\times P_{0,{\rm fid}}(q_1)\,\delta P_0(q_2),
\\
&\delta\overline{\Gamma}^{(2)}_{\rm 1\mbox{-}loop}(\bfk_1,\bfk_2) =
\int\frac{dq}{2\pi^2}\,K(q;k_1,k_2,k_3)\,\delta P_0(q)
\end{align}
with the angle-averaged kernels $f$, $J$ and $K$ defined by 
\begin{align}
& f(q;k) =3\int\frac{d^2\bfOmg_q}{4\pi}\,F_{\rm sym}^{(3)}(\bfq,-\bfq,\bfk),
\\
& J(q_1,q_2;k) =
15\,\int\frac{d^2\bfOmg_{q_1}\,d^2\bfOmg_{q_2}}{(4\pi)^2}\,
\nonumber\\
&\qquad\qquad\qquad\qquad
\times F_{\rm sym}^{(5)}(\bfq_1,-\bfq_1,\bfq_2,-\bfq_2,\bfk),
\\
&K(q;k_1,k_2,k_3) = 
6\,\int\frac{d^2\bfOmg_q}{4\pi}\,F_{\rm sym}^{(4)}(\bfq,-\bfq,\bfk_1,\bfk_2).
\end{align}
Note that $\bfk_1+\bfk_2=\bfk_3$.

The expression for the correction $P_{\rm corr}$ given above 
contains many integrals 
involving the perturbed linear power spectrum, $\delta P_0$, and some 
of these require  multi-dimensional integrals. 
However, those multi-dimensional integration are separately treated, and 
can be effectively reduced to the one-dimensional integrals as follows,
\begin{widetext}
\begin{align}
&\delta\overline{\Gamma}_{\rm 1\mbox{-}loop}^{(1)}(k)
=\int\frac{dq q^2}{2\pi^2} L^{(1)}(q,k)\,\delta P_0(q),
\\
&\delta\overline{\Gamma}_{\rm 2\mbox{-}loop}^{(1)}(k)
=2\int\frac{dq q^2}{2\pi^2} M^{(1)}(q,k)\,\delta P_0(q),
\\
\nonumber
\\
&\int\frac{d^3\bfq d^3\bfq}{(2\pi)^3}\left[
\Gamma_{\rm reg}^{(2)}(\bfq,\bfk-\bfq;\eta)\right]^2P_{0,{\rm fid}}(|\bfk-\bfq|) \,
\delta P_0(q)=e^{4\eta}\Bigl[
\left(1+\alpha_k\right)^2\int\frac{dq\,q^2}{2\pi^2}
X^{(2)}(q,k)\,\delta P_0(q)
\nonumber\\
&\qquad\qquad+2e^{2\eta}\,
\left(1+\alpha_k\right)\int\frac{dq\,q^2}{2\pi^2}
Y^{(2)}(q,k)\,\delta P_0(q) +e^{4\eta}\,
\int\frac{dq\,q^2}{2\pi^2}Z^{(2)}(q,k)\,\delta P_0(q)
\Bigr]\exp\left\{-2\alpha_k\right\},
\\
\nonumber
\\
&\int \frac{d^3\bfq}{(2\pi)^3}\,\Gamma_{\rm reg}^{(2)}(\bfq,\bfk-\bfq;\eta)
\,
\delta \Gamma_{\rm reg}^{(2)}(\bfq,\bfk-\bfq;\eta)\,
P_{0,{\rm fid}}(|\bfk-\bfq|) P_{0,{\rm fid}}(q)
\nonumber\\
&\qquad=
e^{6\eta}\Bigl[
(1+\alpha_k)\,\int\frac{dpp^2}{2\pi^2}\,
Q^{(2)}(p,k)\,\delta P_0(p)+
e^{2\eta}\,\int\frac{dpp^2}{2\pi^2}\,
R^{(2)}(p,k)\,\delta P_0(p)\Bigr]e^{-2\alpha_k}, 
\\
\nonumber
\\
&\int\frac{d^3\bfq d^3\bfq}{(2\pi)^6}\left[
\Gamma_{\rm reg}^{(3)}(\bfp,\bfq,\bfk-\bfp-bfq;\eta)\right]^2\,
P_{0,{\rm fid}}(|\bfk-\bfp-\bfq|) P_{0,{\rm fid}}(p)\,
\delta P_0(q)=e^{6\eta}\,
e^{-2\alpha_k}\,\int\frac{dq\,q^2}{2\pi^2}\,
S^{(3)}(q,k)\,\delta P_0(q).
\end{align}
In the above, the kernels of the integrals, $L$, $M$, $X$, $Y$, 
$Z$, $Q$, $R$, and $S$, additionally need to be computed, but 
we only have to evaluate them once for each fiducial cosmological model, 
\begin{align}
&L^{(1)}(q,k)=f(q;k),
\label{eq:L1}\\
&M^{(1)}(q,k)=\int\frac{dp\,p^2}{2\pi^2} J(p,q;k)\,P_{0,{\rm fid}}(p) ,
\label{eq:M1}\\
&X^{(2)}(q,k)=\frac{1}{2}\int_{-1}^1d\mu_q\,
\left[F^{(2)}_{\rm sym}(\bfq,\bfk-\bfq)\right]^2\,
P_{0,{\rm fid}}(\sqrt{k^2-2kq\,\mu_q+q^2}),
\label{eq:X2}\\
&Y^{(2)}(q,k)=\frac{1}{2}\int_{-1}^1d\mu_q\,
F^{(2)}_{\rm sym}(\bfq,\bfk-\bfq)\,\overline{\Gamma}^{(2)}_{\rm 1\mbox{-}loop}
(\bfq,\bfk-\bfq)\,
P_{0,{\rm fid}}(\sqrt{k^2-2kq\,\mu_q+q^2}),
\label{eq:Y2}\\
&Z^{(2)}(q,k)=\frac{1}{2}\int_{-1}^1d\mu_q\,
\left[\overline{\Gamma}^{(2)}_{\rm 1\mbox{-}loop}
(\bfq,\bfk-\bfq)\right]^2\,
P_{0,{\rm fid}}(\sqrt{k^2-2kq\,\mu_q+q^2}),
\label{eq:Z2}\\
&Q^{(2)}(p,k)=\int\frac{d^3\bfq}{(2\pi)^3}
F^{(2)}_{\rm sym}(\bfq,\bfk-\bfq)\,K(p:q,|\bfk-\bfq|,k)\,
P_{0,{\rm fid}}(|\bfk-\bfq|)\,P_{0,{\rm fid}}(q),
\label{eq:Q2}\\
&R^{(2)}(p,k)=\int\frac{d^3\bfq}{(2\pi)^3}
\overline{\Gamma}^{(2)}_{\rm 1\mbox{-}loop}(\bfq,\bfk-\bfq)\,
K(p;q,|\bfk-\bfq|,k)\,
P_{0,{\rm fid}}(|\bfk-\bfq|)\,P_{0,{\rm fid}}(q),
\label{eq:R2}\\
&S^{(3)}(q,k)=\frac{1}{2}\int_{-1}^1 d\mu_q\,\int\frac{d^3\bfp}{(2\pi)^3}
\left[F^{(3)}_{\rm sym}(\bfp,\bfq,\bfk-\bfp-\bfq)\right]^2\,
P_{0,{\rm fid}}(|\bfk-\bfp-\bfq|)\,P_{0,{\rm fid}}(p),
\label{eq:S3}
\end{align}
\end{widetext}
with the variable $\mu_q$ defined by $\mu_q=(\bfk\cdot\bfq)/(k\,q)$.

Note that similar to the one-loop case, the correction at two-loop 
order also possesses a one-parameter degree of freedom corresponding to a 
re-scaling the power spectrum amplitude of fiducial model, 
$P_{0,{\rm fid}}\to c\,P_{0,{\rm fid}}$. The power spectrum difference 
$\delta P_0$ can then be made small securing the rapid convergence of this 
expansion.

Finally, the data set of kernel functions given above are supplemented 
in the \RegPT~code, with $301\times301$ 
logarithmic arrays in $(k,q)$ space.  
For specific three fiducial models (i.e., 
{\tt wmap3}, {\tt M001}, and {\tt M023}), 
the data have been obtained using the method of Gaussian quadrature up to 
three-dimensional integrals and Monte Carlo technique for four-dimensional
integral. Together with un-perturbed part of the PT corrections, 
these can be used as fast calculations of power spectrum at two-loop order.

\section{Code description}
\label{appendix:RegPT_code}

In this Appendix, we present a detailed description of the fortran 
code, \RegPT, which computes the power spectrum 
and correlation function of density fields valid at weakly non-linear regime 
of gravitational clustering. 

\subsection{Overview}
\label{subsec:overview}

The code, \RegPT, is compiled with the fortran compilers, {\tt ifort} or
{\tt gfortran}. It computes
the power spectrum in flat wCDM class models 
based on the \RegPT~treatment
when provided with either of 
transfer function or matter power spectrum. 
It then gives the multiple-redshift outputs for power spectrum, 
and optionally provides correlation function data. 
We have implemented two major options for power spectrum calculations: 

\begin{itemize}
\item {\tt -fast}: Applying the reconstruction method described in 
Sec.~\ref{subsec:PTreconst_1-loop}, this option quickly computes 
the power spectrum at two-loop level (typically a few seconds), using   
the pre-computed data set of PT kernels for fiducial cosmological 
models. We provide the data set for three fiducial models ({\tt wmap3}, 
{\tt M001}, and {\tt M023}, see Table~\ref{tab:fiducial_model}), and the code 
automatically finds an appropriate fiducial model to 
closely match the result of rigorous PT calculation with direct-mode. 

\item {\tt -direct}: With this option,  the code first applies the fast 
method, and then follows the regularized 
expression for power spectrum (see Eq.[\ref{eq:pk_Gamma_reg_2loop}] with 
regularized propagators [\ref{eq:Gamma1_reg}]-[\ref{eq:Gamma3_reg}]) to 
directly evaluate the multi-dimensional integrals (it typically takes a few 
minutes). The output results are the power spectrum of direct calculation 
and difference of the results between fast and direct method. Further, 
the code gives the data set of PT diagrams necessary for power spectrum 
calculations, from which we can construct the power spectrum. 
We provide a supplemental code, {\tt read\_stfile.f}, with which the power 
spectrum and correlation function can be evaluated from the diagram data set 
in several PT methods, including the standard PT 
and Lagrangian resummation theory (LRT) \cite{Matsubara:2007wj,Okamura:2011nu} 
as well as RegPT treatment (see Appendix \ref{subsubsec:diagram_data}). 
\end{itemize}

In addition, the code supports the option, {\tt -direct1loop}, to compute 
the power spectrum at one-loop order. Although this is based on the direct 
calculation with multi-dimensional integration (see 
Eq.~[\ref{eq:pk_Gamma_reg_1loop}] with regularized propagators 
[\ref{eq:Gamma1_reg_1loop}][\ref{eq:Gamma2_reg_tree}]), 
the one-loop expression involves two-dimensional integrals at most, 
and thus the computational cost is less expensive. It is potentially useful
for the computation of high-$z$  correlation function and power spectrum.

\bigskip

\subsection{Setup}
\label{subsec:setup}
The \RegPT~code is available at
\begin{widetext}
\verb|              http://www-utap.phys.s.u-tokyo.ac.jp/~ataruya/regpt_code.html|
\vspace*{0.5cm}
\end{widetext}
A part of \RegPT~code uses the library for Monte Carlo integration, 
CUBA \cite{Hahn:2004fe}. 
Before compiling the codes, users should download the library package 
{\tt cuba-1.5}, and correctly build the file, {\tt libcuba.a}, 
compatible with the architecture of user's platform. This can be done in 
the directory {\tt /Cuba-1.5}, and just type {\tt ./configure} 
and {\tt make lib}. After placing the library file {\tt libcuba.a} 
at the directory {\tt /RegPT/src}, 
users can use the {\tt Makefile} to create the main executable file, 
{\tt RegPT.exe}. Note that currently available compilers are 
intel fortran compiler, {\tt ifort}, and GNU fortran compiler, {\tt gfortran}.

\subsection{Running the code}
\label{subsec:running_code}

Provided with linear power spectrum or transfer function data, the code runs 
with a set of options, and computes power spectrum. 
Users can specify the options in the command line, or using the parameter file 
(suffix of file name should be {\tt .ini}). Sample of parameter file is 
supplied in the code (see directory {\tt /RegPT/example}). 

For running the code with the command-line options, a simple example 
is (assuming the code is placed at the directory, {\tt /RegPT})

\begin{widetext}
\begin{align}
\mbox{\tt  ./RegPT.exe -spectrum -infile matterpower\_wmap5.dat -nz 2 0.5 1.0 }
\nonumber
\end{align}
\end{widetext}

In the above example, 
the code first reads the input data file, {\tt matterpower\_wmap5.dat}, 
which is assumed 
to contain linear power spectrum data consisting of two columns,  
i.e., $k$ and $P_0(k)$. By default setting, {\tt fast} mode is chosen, 
and the output result of power spectrum is saved to 
{\tt pk\_RegPT.dat}. With the option {\tt -nz 2 0.5 1.0}, the 
output file contains the power spectrum results at two redshifts, 
$z=0.5$ and $1.0$ (see Appendix \ref{subsubsec:power_spectrum_data} for 
output format). 
Note that by default, the code adopts specific values of cosmological 
parameters. Making use of options, users can change the value of 
cosmological parameters appropriately, consistently with input power 
spectrum (or transfer function) data.

Here we summarize the available options to run the code: 

\begin{itemize}
\item Verbose level for output message

\vspace*{0.3cm}
\noindent
{\tt -verbose}\,\,\,$n$: This sets the verbose level for output information on the progress of numerical computation. The available level $n$ is 1 or 2 (default: {\tt -verbose 1}). 

\vspace*{0.1cm}

\noindent
{\tt -noverbose}: This option suppresses the message while running the code.

\vspace*{0.3cm}

\item Input data file

\vspace*{0.3cm}
\noindent
{\tt -infile}\,\,\,[\,file\,]: Input file name of power spectrum or transfer 
    function data is specified (default: {\tt -infile matterpower.dat}). 

\vspace*{0.1cm}

\noindent
{\tt -path}\,\,\,[\,path to input file\,]: This specifies the path to the input file 
    (default: {\tt -path ./}).  

\vspace*{0.1cm}

\noindent
{\tt -spectrum}: With this option, the code assumes that the input 
    file is power spectrum data. The data consists of two columns, i.e., 
    wavenumber (in units of $h\,$Mpc$^{-1}$) and matter power spectrum 
    (in units of $h^{-3}$\,Mpc$^{3}$) (default: {\tt -spectrum}). The 
    normalization of power spectrum amplitude can be made with the option 
    {\tt -sigma8}. 

\vspace*{0.1cm}

\noindent
{\tt -transfer}: With this option, the code assumes that the input 
    file is the transfer function data created by CAMB. The data should 
    contain $7$ columns, among which the code uses the first and seven columns 
    (wavenumber in units of $h\,$Mpc$^{-1}$ and matter transfer function). 
    The normalization of power spectrum amplitude can be made with either of
    the option {\tt -sigma8} or {\tt -samp} and {\tt -spivot}. 

\vspace*{0.3cm}

\item Specification of cosmological parameters

\vspace*{0.3cm}

\noindent
{\tt -sigma8}\,\,\,$\sigma_8$: 
This option sets the power spectrum normalization by $\sigma_8$ 
(default: {\tt -sigma8 0.817}). For $\sigma_8<0$, the code will skip 
the $\sigma_8$ normalization. 

\noindent
{\tt -samp}\,\,\,$A_{\rm s}$: 
This option sets the amplitude of power spectrum at pivot 
scale $k_{\rm pivot}$ (default: {\tt -samp 2.1e-9}). This option is used 
for normalization of transfer function data, and is valid when 
the option {\tt -transfer} is specified. 

\vspace*{0.1cm}

\noindent
{\tt -spivot}\,\,\,$k_{\rm pivot}$: 
This option sets the pivot scale of CMB normalization in units of 
Mpc$^{-1}$ (default: {\tt -spivot 0.05}). This option is used for 
normalization of transfer function data, and is valid when 
the option {\tt -transfer} is specified. 

\vspace*{0.1cm}

\noindent
{\tt -omegam}\,\,\,$\Omega_{\rm m}$: This option sets the mass density parameter 
(default: {\tt -omegam 0.279}). This is used to estimate the linear 
    growth factor and to compute the smooth reference spectrum, 
    $P_{\rm no\mbox{-}wiggle}(k)$.

\vspace*{0.1cm}

\noindent
{\tt -omegab}\,\,\,$\Omega_{\rm b}$: This option sets the baryon density 
parameter (default: {\tt -omegab 0.165*omegam}). This is used to 
    compute the smooth reference spectrum, $P_{\rm no\mbox{-}wiggle}(k)$.

\vspace*{0.1cm}

\noindent
{\tt -ns}\,\,\,$n_{\rm s}$: This option sets the scalar spectral index. 
    This is used to compute the linear power spectrum from the 
    transfer function data (option {\tt -transfer} should be specified), 
    and to compute the smooth reference spectrum, $P_{\rm no\mbox{-}wiggle}(k)$.

\vspace*{0.1cm}

\noindent
{\tt -w}\,\,\,$w$: This option sets the equation of state for dark energy 
(default: {\tt -w -1.0}). This is used to estimate the linear growth 
factor. 

\vspace*{0.1cm}

\noindent
{\tt -h}\,\,\,$h$: This option sets the Hubble parameter 
(default: {\tt -h 0.701}). This is used to compute the power spectrum from 
the transfer function data, and to compute the smooth reference 
spectrum, $P_{\rm no\mbox{-}wiggle}(k)$.

\vspace*{0.1cm}

\noindent
{\tt -camb}\,\,\,[\,output parameter file of camb\,]: With this option, 
the code 
reads the CAMB output parameter file, and specifies the cosmological parameters 
    ($\Omega_{\rm m}$, $\Omega_{\rm b}$, $w$, $h$, $n_{\rm s}$, $A_{\rm s}$, 
    $k_{\rm pivot}$). 

\vspace*{0.3cm}

\item Calculation mode of \RegPT

\vspace*{0.3cm}

\noindent
{\tt -fast}: This option adopts the fast method of power 
    spectrum calculation to give \RegPT~results. This is default setting.

\vspace*{0.1cm}

\noindent
{\tt -direct}: This option first applies the fast method, and then 
    follow the direct method for \RegPT~calculation.

\vspace*{0.1cm}

\noindent
{\tt -direct1loop}: With this option, the code adopts direct method 
    to compute the power spectrum at one-loop order.  

\vspace*{0.3cm}

\item Setup of fiducial models for fast- and direct-mode calculations

\vspace*{0.3cm}

\noindent
{\tt -datapath}\,\,\,[\,path to data directory\,]: This option specifies 
    the path to the data files used for power spectrum calculation with 
    fast and direct methods 
    (default: {\tt -datapath data/}). In the directory specified with this 
    option, the data set of kernel functions given in Appendix 
    \ref{sec:PTreconst_2-loop} and 
    un-perturbed part of power spectrum corrections, as well as the 
    matter power spectrum should be stored for three fiducial 
    cosmological models ({\tt wmap3}, {\tt M001}, {\tt M023}). 
    
\vspace*{0.1cm}

\noindent
{\tt -fiducial}\,\,\,[\,model\,]: This option sets the specific fiducial 
    model among the three, {\tt wmap3}, {\tt M001}, and {\tt M023} (in 
    default setting, the code automatically selects an appropriate fiducial 
    model). 

\vspace*{0.3cm}

\item Output data file

\vspace*{0.3cm}

\noindent
{\tt -xicompute}: With this option, the code computes the correlation 
    function after power spectrum calculations, and creates the output file. 

\vspace*{0.1cm}

\noindent
{\tt -nz}\,\,\,$n$\,\,$z_1\,\,\cdots\,\,z_n$: This option specifies the output 
redshifts for power spectrum calculations. The integer $n$ specifies the 
number of redshifts, and subsequent arguments specify the value of each 
redshift (default: {\tt -nz 1 1.0}). 

\vspace*{0.1cm}

\noindent
{\tt -pkfile}\,\,\,[\,file\,]: This option sets the output file name of power 
spectrum data (default: {\tt pk\_RegPT.dat}). 

\vspace*{0.1cm}

\noindent
{\tt -xifile}\,\,\,[\,file\,]: This option sets the output file name of 
correlation function data (default: {\tt xi\_RegPT.dat}). 

\vspace*{0.1cm}

\noindent
{\tt -stfile}\,\,\,[\,file\,]: This option sets the output file name of 
PT diagram data (default: {\tt st\_PT.dat}). 

\end{itemize}

\subsection{Output file format}
\label{subsec:output_files}

In what follows, wavenumber $k$ and separation $r$ are in units of 
$h\,$Mpc$^{-1}$ and $h^{-1}\,$Mpc, respectively. 
All the power spectrum data are assumed to 
be in units of $h^{-3}\,$Mpc$^3$. 

\subsubsection{Power spectrum data}
\label{subsubsec:power_spectrum_data}

By default, \RegPT~code creates the output file for the power spectrum data 
(default file name is {\tt pk\_RegPT.dat}). The columns of this file include
\begin{align}
k,\,\, \mbox{[data for $z_1$]},\,\, \mbox{[data for $z_2$]},\,\, \cdots, 
\mbox{[data for $z_n$]}
\nonumber
\end{align}
The first column is the wavenumber, while the bracket 
$\mbox{[data for $z_i$]}$ represents a set of power spectra 
at given redshift $z_i$ and wavenumber $k$. Number of the data set
is specified with the option {\tt -nz}, and each data contains
\begin{align}
P_{\rm no\mbox{-}wiggle}(k,z_i),\,\, 
P_{\rm lin}(k,z_i),\,\, P_{\rm RegPT}(k,z_i),\,\, \mbox{Err}(k)
\nonumber
\end{align}
Here, the spectrum $P_{\rm no\mbox{-}wiggle}$ is the smooth reference spectrum 
calculated  
from the no-wiggle formula of linear transfer function in 
Ref.~\cite{Eisenstein:1997ik}, $P_{\rm lin}$ is the linearly extrapolated 
spectrum, and $P_{\rm RegPT}(k,z_i)$ represents the power spectrum 
based on the RegPT 
calculations with fast and/or direct method 
(depending on the choice of options, 
{\tt -fast}, {\tt -direct} or {\tt -direct1loop}). 
The last column, $\mbox{Err}$, 
usually sets to zero, but with the option {\tt -direct}, 
it gives the difference of the power spectra between fast and direct methods.

\subsubsection{Correlation function data}
\label{subsubsec:correlation_function_data}

With the option {\tt -xicompute}, the code also provides the output file for 
correlation function data (default file name is {\tt xi\_RegPT.dat}). Similar 
to the power spectrum data, the structure of the data is 
\begin{align}
r,\,\, \mbox{[data for $z_1$]},\,\, \mbox{[data for $z_2$]},\,\, \cdots, 
\mbox{[data for $z_n$]}
\nonumber
\end{align}
The first column is the separation, while the bracket 
$\mbox{[data for $z_i$]}$ represents a set of correlation functions
given at redshift $z_i$ and separation $r$, containing two columns: 
\begin{align}
\xi_{\rm lin}(r,z_i),\,\, \xi_{\rm RegPT}(r,z_i) 
\nonumber
\end{align}
These are simply obtained from the output results of power spectrum based on
the expression (\ref{eq:def_xi}). Note that 
the range of wavenumber for output power spectrum is 
restricted to the wavenumber coverage of input linear spectrum 
(or transfer function). To get a convergent result of correlation 
functions, users may have to supply the input data file with 
a sufficiently wide range of wavenumber 
(e.g., $10^{-3}\leq k\leq 10$\,$h\,$Mpc$^{-1}$).

\subsubsection{Diagram data}
\label{subsubsec:diagram_data}

When users specifies the {\tt -direct} option, the code additionally 
provides a set of PT diagram data necessary for power spectrum computation,  
from which we can construct the power spectrum at one- and two-loop order. 
The output file (default file name is {\tt st\_PT.dat}) includes the following 
columns:
\begin{widetext}
\begin{align}
k,\,\,P_{\rm no\mbox{-}wiggle}(k),\,\, P_{\rm lin}(k),\,\, 
\overline\Gamma^{(1)}_{1\mbox{-}{\rm loop}}(k), \,\,
\overline\Gamma^{(1)}_{2\mbox{-}{\rm loop}}(k), \,\,
P_{\rm corr}^{(2){\rm tree\mbox{-}tree}}(k), \,\,
P_{\rm corr}^{(2){\rm tree\mbox{-}1loop}}(k), \,\,
P_{\rm corr}^{(2){\rm 1loop\mbox{-}1loop}}(k),\,\,
P_{\rm corr}^{(3){\rm tree\mbox{-}tree}}(k)
\nonumber
\end{align}
\end{widetext}
Here, the power spectra $P_{\rm no\mbox{-}wiggle}$ and $P_{\rm lin}$ 
are basically the same data as contained in the power spectrum file, 
but these are the extrapolated data at $z=0$ (that is, $P_{\rm lin}$ 
corresponds to $P_0$). The function 
$\overline\Gamma^{(1)}_{n\mbox{-}{\rm loop}}$ is the two-point propagator 
of the standard PT expansion (see definition [\ref{eq:Gamma-p_nloop}]). 
The functions in the remaining four columns, $P_{\rm corr}^{(2){\rm tree\mbox{-}tree}}$, 
$P_{\rm corr}^{(2){\rm tree\mbox{-}1loop}}$, 
$P_{\rm corr}^{(2){\rm 1loop\mbox{-}1loop}}$, and $P_{\rm corr}^{(3){\rm tree\mbox{-}tree}}$, 
are defined by
\begin{widetext}
\begin{align}
&P^{(2)\rm tree\mbox{-}tree}_{\rm corr}(k) = 
2\int\frac{d^3\bfq}{(2\pi)^3}F_{\rm sym}^{(2)}(\bfq,\bfk-\bfq)
F_{\rm sym}^{(2)}(\bfq,\bfk-\bfq)
P_0(q)P_0(|\bfk-\bfq|),
\label{eq:P2_tree-tree}
\\
&P^{(2)\rm tree\mbox{-}1loop}_{\rm corr}(k) = 
4\int\frac{d^3\bfq}{(2\pi)^3} 
F_{\rm sym}^{(2)}(\bfq,\bfk-\bfq)
\overline{\Gamma}_{\rm 1\mbox{-}loop}^{(2)}(\bfq,\bfk-\bfq)
P_0(q)P_0(|\bfk-\bfq|),
\label{eq:P2_tree-1loop}
\\
&P^{(2)\rm 1loop\mbox{-}1loop}_{\rm corr}(k) = 
2\int\frac{d^3\bfq}{(2\pi)^3} 
\overline{\Gamma}_{\rm 1\mbox{-}loop}^{(2)}(\bfq,\bfk-\bfq)
\overline{\Gamma}_{\rm 1\mbox{-}loop}^{(2)}(\bfq,\bfk-\bfq)
P_0(q)P_0(|\bfk-\bfq|),
\label{eq:P2_1loop-1loop}
\\
&P^{(3)\rm tree\mbox{-}tree}_{\rm corr}(k) = 
6\int\frac{d^3\bfp\,d^3\bfq}{(2\pi)^6} 
F^{(3)}_{\rm sym}(\bfp, \bfq,\bfk-\bfp-\bfq)
F^{(3)}_{\rm sym}(\bfp, \bfq,\bfk-\bfp-\bfq)
P_0(p)P_0(q)P_0(|\bfk-\bfp-\bfq|).
\label{eq:P3_tree-tree}
\end{align}
\end{widetext}

Provided the data set above, the power spectrum can be constructed 
with 
\begin{widetext}
\begin{align}
&P_{\rm 1\mbox{-}loop}^{\rm RegPT}(k;\eta)=e^{2\eta}\,e^{-2\alpha_k}\left[
\left\{1+\alpha_k+e^{2\eta}\,\overline{\Gamma}_{\rm 1\mbox{-}loop}^{(1)}(k)\right\}^2\,
P_0(k)+e^{2\eta}\,P^{(2)\rm tree\mbox{-}tree}_{\rm corr}(k)\right],
\label{eq:pk_RegPT-1loop}
\\
&P_{\rm 2\mbox{-}loop}^{\rm RegPT}(k;\eta)=e^{2\eta}\,e^{-2\alpha_k}
\Bigl[\,\,
\Bigl\{1+\alpha_k+\frac{\alpha_k^2}{2}
+e^{2\eta}\,\overline{\Gamma}_{\rm 1\mbox{-}loop}^{(1)}(k)\left(1+\alpha_k\right)
+e^{4\eta}\,\overline{\Gamma}_{\rm 2\mbox{-}loop}^{(1)}(k)\Bigr\}^2
\,P_0(k)
\nonumber\\
&\quad
+e^{2\eta}\left\{(1+\alpha_k)^2P^{(2)\rm tree\mbox{-}tree}_{\rm corr}(k)
+e^{2\eta}(1+\alpha_k)P^{(2)\rm tree\mbox{-}1loop}_{\rm corr}(k)
+e^{4\eta}\,P^{(2)\rm 1loop\mbox{-}1loop}_{\rm corr}(k)\right\}
+e^{4\eta}\,P^{(3)\rm 1loop\mbox{-}1loop}_{\rm corr}(k)\,\,\Bigr],
\label{eq:pk_RegPT-2loop}
\end{align}
\end{widetext}
for the \RegPT~calculation at one- and two-loop order, respectively. Here, 
$\alpha_k$ is given by $\alpha_k=k^2\sigmav^2e^{2\eta}/2$ 
with $\sigmav$ being the dispersion of displacement field 
(see Eq.~[\ref{eq:sigam_v}]).   Note that the diagram data set can be 
also used to compute the power spectrum in the standard PT calculations:  
\begin{widetext}
\begin{align}
&P_{\rm 1\mbox{-}loop}^{\rm SPT}(k;\eta)=e^{2\eta}\,P_0(k)+ e^{4\eta}\left[2\,P_0(k)\,
\overline{\Gamma}_{1\mbox{-}{\rm loop}}^{(1)}(k) +
P^{(2)\rm tree\mbox{-}tree}_{\rm corr}(k)\right],
\label{eq:pk_SPT_1loop}
\\
&P_{\rm 2\mbox{-}loop}^{\rm SPT}(k;\eta)=P_{\rm 1\mbox{-}loop}^{\rm SPT}(k;\eta)+ 
e^{6\eta}\,\Bigl[\,\,
P_0(k)\,\{\overline{\Gamma}_{1\mbox{-}{\rm loop}}^{(1)}(k)\}^2+
P^{(3)\rm tree\mbox{-}tree}_{\rm corr}(k)+ P^{(2)\rm tree\mbox{-}1loop}_{\rm corr}(k)
+2P_0(k)\,\overline{\Gamma}_{2\mbox{-}{\rm loop}}^{(1)}(k)
\Bigr]. 
\label{eq:pk_SPT_2loop}
\end{align}
\end{widetext}

With the supplemental code, {\tt read\_stfile.f}, users can easily compute 
the power spectrum in both \RegPT~and standard PT treatments. 
The code also provides the power spectrum result for 
LRT \cite{Matsubara:2007wj,Okamura:2011nu}. 
A brief instruction on how to run the code and the output format of data
is described in the header of the code.

\subsection{Limitation}
\label{subsec:limitation}

Since the \RegPT~code is the PT-based calculation code valid at weakly 
non-linear scales, the applicability of the output results 
is restricted to a certain range of wavenumber in power spectrum. 
We provide an empirical estimate of critical wavenumber $k_{\rm crit}$, 
below which the \RegPT~results are reliable and their accuracy 
can reach a percent level. This is based on Eq.~(\ref{eq:k_crit}) 
with constant value $C=0.7\,(0.3)$ for two-loop (one-loop) 
(see Sec.~\ref{subsubsec:emulator}). With the option {\tt -verbose 2}, 
the code displays the critical wavenumbers at output redshifts. Note 
that the value $k_{\rm crit}$ given here is just a crude estimate, 
and the actual domain of applicability may be somewhat wider or 
narrower. Users should use the output results with a great care.

\bibliographystyle{apsrev}


\end{document}